\documentclass[jmp,aps,showpacs,reprint]{revtex4-1}
\usepackage{graphicx}
\usepackage{dcolumn}
\usepackage{amsmath}
\usepackage{xcolor}
\usepackage{epsfig}
\RequirePackage{xspace}
\usepackage{threeparttable}

\usepackage{relsize}

\begin{document}

\title{
\large \bfseries \boldmath Study of $B\to PP$ decays in the modified perturbative QCD approach}

\author{Sheng L\"{u}} \email{shenglyu@mail.nankai.edu.cn}
\author{Ru-Xuan Wang} \email{wangrx@mail.nankai.edu.cn}
\author{Mao-Zhi Yang}\email{yangmz@nankai.edu.cn}
\affiliation{School of Physics, Nankai University, Tianjin 300071, People's Republic of China}

\date{\today}


\begin{abstract}
We study the nonleptonic decays of $B\to PP$ in the modified perturbative QCD approach, where $P$ stands for pseudoscalar mesons. Transverse momenta of partons and the Sudakov factor are included, which help to suppress the contributions of soft interactions. The wave function of the $B$ meson obtained from the relativistic potential model is used, and then the contributions in the infrared region cannot be suppressed completely. So a soft cutoff scale and soft form factors are introduced. The contributions with the scale higher than the soft cutoff scale are calculated with perturbative QCD, while the contributions lower than the cutoff scale are replaced by the soft form factors. To explain experimental data, we find that contributions of color-octet operators for the quark-antiquarks in the mesons in the final state need to be considered. The contributions of the color-octet operators are parametrized by a few parameters with the help of SU(3) flavor symmetry and symmetry breaking. These parameters for color-octet contributions are universal for all the non-leptonic decay modes of the $B$ meson, where the mesons in the final state belong to the same flavor SU(3) nonet. Both the branching ratios and $CP$ violations are studied. We find that the theoretical calculation can well explain the experimental data of $B$ factories.
\end{abstract}

\maketitle
\section{Introduction}

$B$ meson decays are important for studying the mechanism of electroweak and strong interactions in particle decays. More precision experimental data have been collected by $B$ factories and LHCb experiments for the last two decades \cite{PDG2022}, which greatly helped the development of the theoretical methods for calculating $B$ decays. One of the difficulties in this area is how to treat the QCD effects in $B$ decays. Several methods have been developed to calculate the effect of strong interaction in QCD on the bases of the factorization theorem, which are the perturbative QCD (PQCD) approach \cite{PQCD1,PQCD2,PQCD3}, QCD factorization (QCDF) approach \cite{QCDf1,QCDf2,QCDf3,QCDf4}, and the soft collinear effective theory \cite{SCET1,SCET2,SCET3,SCET4,SCET5,SCET6}. By confronting theoretical predictions to experimental data, it is found that the predictions for most decay modes of the $B$ meson are consistent with experimental measurements, which illustrates the successful aspect of these theoretical methods for treating $B$ decays. But a few serious problems emerged, such as the $B\to \pi\pi$ and $K\pi$ puzzles, which is when the branching ratio of $B\to \pi^0\pi^0$ measured by experiment is several times larger than theoretical predictions, while the theoretical predictions for the other modes of $B\to \pi\pi$ decays are approximately consistent with experimental data. For the $B\to K\pi$ decays, the measured direct $CP$ asymmetries of $B^{\pm}\to \pi^0 K^{\pm}$ and $B^0\to \pi^{\mp} K^{\pm}$ are dramatically different, which apparently deviate from theoretical expectation; see Refs. \cite{LiMiSa2005,Li-Mishima2011,Li-Mishima2014}. The measured branching ratios of $B\to K\pi$ decays are also puzzling from the theoretical point of view \cite{BFRS2003}.

Many efforts have been made to solve the $B\to \pi\pi$ and $K\pi$ puzzles within \cite{LiMiSa2005,Li-Mishima2011,Li-Mishima2014,bai2014revi,bai2014revi,LLX2016,xiao2022,
cheng-chua2005,cheng-chua2009a,cheng-chua2009b,CSYL2014,chua2018} and beyond \cite{Bar2004,Bae2005,Arn2006,Kim2008,Bea2018,Dat2019} the standard model. Progress has been made in understanding the $\pi\pi$ and $K\pi$ puzzles by these efforts. Tensions between experimental data and theoretical calculations are diminished, but other ways investigating such problems are still welcome.

In Refs. \cite{Lu-Yang2021,lu-yang2023,wang-yang2023}, the wave function of the $B$ meson solved from the QCD-inspired relativistic potential model \cite{Yang2012,LY2014,LY2015,SY2017,SY2019} is used in the PQCD approach. We find that the long-distance contribution cannot be suppressed by the Sudakov factor effectively with the new $B$ wave function being used. So a cutoff scale $\mu_c$ that separates the soft and hard contributions in QCD has been introduced. The contribution with the scale higher than $\mu_c$ can be calculated with PQCD method, while the contribution with scale lower than $\mu_c$ should be replaced by soft form factors. By confronting the theoretical calculations to the experimental data on $B\to\pi\pi$ and $K\pi$ decays, we find that the contributions of the quark-antiquark pairs in color-octet states that form the final mesons in the long-distance region need to be introduced to explain the experimental data. By taking reasonable values for the parameters for the color-octet contributions, the $\pi\pi$ and $K\pi$ puzzles can be resolved with the modified PQCD approach.

In this work, we extend our previous works in $B\to \pi\pi$ and $K\pi$ decays \cite{lu-yang2023,wang-yang2023} to more decay modes for the $B$ meson, where two pseudoscalar mesons are included in the final state. Compared with our previous works in Refs. \cite{lu-yang2023,wang-yang2023}, further progress is made: (1) More decay modes of $B\to PP$ decays are studied in the modified PQCD approach. Both the branching ratios and $CP$ asymmetries are calculated and compared with experimental data. (2) The parameters for the color-octet contributions and production form factor are treated by considering SU(3) flavor symmetry and the symmetry breaking. Therefore, these parameters are no longer directly final-state-dependent parameters. They are universal for the mesons in the same SU(3) flavor nonet, which makes our method with prediction power. By selecting reasonable values for these parameters, we find that the theoretical results can be consistent with experimental data.

The paper is organized as follows. The perturbative part including the leading and next-to-leading-order (NLO) contributions in QCD is presented in Sec. II. The soft form factors are introduced and their contributions are calculated in Sec. III. The color-octet contributions are discussed in Sec. IV. Section V is devoted to the analysis of the soft parameters under SU(3) flavor symmetry and its breaking. Section VI is for numerical calculation and discussion. A brief summary is given in the final section.

\section{The hard amplitude in perturbative QCD}

\subsection{The effective Hamiltonian}

The effective Hamiltonian for charmless hadronic decays of the $B$ meson induced by the $b\to d (s)$ transition is \cite{Hamiltanion1996}
\begin{eqnarray}
	\mathcal{H}_{\mathrm{eff}} &=& \frac{G_F}{\sqrt{2}}\bigg[ V_{ub}V_{uq}^*\big(C_1O_1^u+C_2 O_2^u\big)\nonumber \\
	&-&V_{tb}V_{tq}^*\bigg(\sum_{i=3}^{10} C_i O_i+C_{8\textsl{g} } O_{8\textsl{g} } \bigg) \bigg],
\end{eqnarray}
where $G_F = 1.16638 \times 10^{-5}~\mathrm{GeV}^{-2}$ is the Fermi constant, $V_{ub}V_{uq}^*$ and $V_{tb}V_{tq}^*$ are the products of the Cabibbo-Kobayashi-Maskawa (CKM) matrix elements with $q=d$ or $s$, $C_i$'s the Wilson coefficients, and the operators are
\begin{eqnarray}
	&&O_1^u = \bar{q}_{\alpha}\gamma^{\mu}L u_{\beta}\cdot \bar{u}_{\beta}\gamma_{\mu}L b_{\alpha},
	 \nonumber  \\
	&&O_2^u = \bar{q}_{\alpha}\gamma^{\mu}L u_{\alpha}\cdot \bar{u}_{\beta}\gamma_{\mu}L b_{\beta},
	\nonumber  \\	
	&&O_3 = \bar{q}_{\alpha}\gamma^{\mu}L b_{\alpha}\cdot \sum_{q'}\bar{q}'_{\beta}\gamma_{\mu}L q'_{\beta},
	\nonumber  \\
	&&O_4 = \bar{q}_{\alpha}\gamma^{\mu}L b_{\beta}\cdot \sum_{q'}\bar{q}'_{\beta}\gamma_{\mu}L q'_{\alpha},
	\nonumber  \\	
	&&O_5 = \bar{q}_{\alpha}\gamma^{\mu}L b_{\alpha}\cdot \sum_{q'}\bar{q}'_{\beta}\gamma_{\mu}R q'_{\beta},
	\nonumber  \\	
	&&O_6 = \bar{q}_{\alpha}\gamma^{\mu}L b_{\beta}\cdot \sum_{q'}\bar{q}'_{\beta}\gamma_{\mu}R q'_{\alpha},	\nonumber  \\
	&&O_7 = \frac{3}{2}\bar{q}_{\alpha}\gamma^{\mu}L b_{\alpha}\cdot \sum_{q'} e_{q'} \bar{q}'_{\beta}\gamma_{\mu}R q'_{\beta},
	\nonumber  	
\end{eqnarray}	
\begin{eqnarray}
	&&O_8 =  \frac{3}{2}\bar{q}_{\alpha}\gamma^{\mu}L b_{\beta}\cdot \sum_{q'} e_{q'} \bar{q}'_{\beta}\gamma_{\mu}R q'_{\alpha},
	\nonumber  \\	
	&&O_9 =  \frac{3}{2}\bar{q}_{\alpha}\gamma^{\mu}L b_{\alpha}\cdot \sum_{q'} e_{q'} \bar{q}'_{\beta}\gamma_{\mu}L q'_{\beta},
	\nonumber\\
  		&&O_{10} =  \frac{3}{2}\bar{q}_{\alpha}\gamma^{\mu}L b_{\beta}\cdot \sum_{q'} e_{q'} \bar{q}'_{\beta}\gamma_{\mu}L q'_{\alpha},  	\nonumber \\
&& O_{8\textsl{g}}=\frac{g_s}{8\pi^2}m_b\bar{q}_\alpha\sigma^{\mu\nu}RT^a_{\alpha\beta}b_\beta G_{\mu\nu}^a, 	\end{eqnarray}
where $\alpha$ and $\beta$ are the color indices, and $L=(1-\gamma_5)$ and $R=(1+\gamma_5)$, which are the left- and right-handed projection operators. The sum relevant to $q'$ runs over all quark flavors being active at $m_b$ scale, that is $q'\in \{u,d,s,c,b\}$.

\subsection{The factorization formula for the decay amplitude and the meson wave functions}

 The momentum transferred by gluons that are exchanged between quarks is generally large in $B$ decays because of the large $b$ quark mass. For the hard dominant region, the decay amplitude of the $B$ meson can be written in a factorized form.  The soft interactions can be absorbed into the meson wave functions. The hard contribution can be calculated perturbatively at quark level. Then the amplitude can be written as
\begin{eqnarray}
&&\mathcal{M}=\int d^3k \int d^3k_1 \int d^3 k_2 \Phi^B(\vec{k},\mu)\nonumber \\
&&\times C(\mu)  H(k,k_1,k_2,\mu) \Phi^{M_1} (k_1,\mu)\Phi^{M_2} (k_2,\mu),
\end{eqnarray}
where $H$ stands for the hard amplitude at quark level, $\Phi^{B,M_1,M_2}$ the meson wave functions, and $C(\mu)$ the Wilson coefficients in the decay.

The spinor wave function of the $B$ meson can be defined by the matrix element $\langle 0| \bar{q}(z)_\beta [z,0]b(0)_\alpha |\bar{B}\rangle$ as
\begin{equation}
\langle 0| \bar{q}(z)_\beta [z,0]b(0)_\alpha |\bar{B}\rangle =\int d^3k \Phi^{B}_{\alpha\beta}(\vec{k})e^{-ik\cdot z},
\end{equation}
where $\Phi^{B}_{\alpha\beta}$ in the right-hand side is the spinor wave function of the $B$ meson, and $[z,0]$ the path-ordered exponential $[z,0]={\cal P}\exp[-ig_sT^a\int_0^1 d\alpha z^\mu A_\mu^a(\alpha z)]$, which is introduced to keep the gauge invariance of the nonlocal quark-antiquark operator.

The spinor wave function $\Phi^{B}_{\alpha\beta}(\vec{k})$ in the $B$ meson rest frame has been derived in Ref. \cite{SY2017} by using the $B$ meson wave function obtained by solving the bound-state equation in the QCD-inspired relativistic potential model in Refs. \cite{Yang2012,LY2014,LY2015}, which is
\begin{eqnarray} \label{B-wave}
  \Phi_{\alpha\beta}(&\vec{k}&)=\frac{-if_Bm_B}{4}K(\vec{k})
\nonumber\\
 && \cdot\Bigg\{(E_Q+m_Q)\frac{1+\not{v}}{2}\Bigg[\Bigg(\frac{k^+}{\sqrt{2}}  +\frac{m_q}{2}\Bigg)\not{n}_+
\nonumber\\
&&+\Bigg(\frac{k^-}{\sqrt{2}}  +\frac{m_q}{2}\Bigg)\not{n}_- -k_{\perp}^{\mu}\gamma_{\mu}  \Bigg]\gamma^5\nonumber\\
&&-(E_q+m_q)\frac{1-\not{v}}{2} \Bigg[  \Bigg(\frac{k^+}{\sqrt{2}}-\frac{m_q}{2}\Bigg)\not{n}_+
\nonumber\\
 && +\Bigg(\frac{k^-}{\sqrt{2}}-\frac{m_q}{2}\Bigg)\not{n}_--k_{\perp}^{\mu}\gamma_{\mu}\Bigg]\gamma^5
  \Bigg\}_{\alpha\beta},\label{eqm}
\end{eqnarray}
where $f_B$ is the decay constant of the $B$ meson, $m_B$ the $B$ meson mass, $E_Q$ and $E_q$ the energies of the heavy and light quarks respectively, and $v$ the four-speed of $B$ meson, i.e., $p_B^\mu=m_B v^\mu$. $n_\pm^\mu$ are two light-like vectors  $n_\pm^\mu=(1,0,0,\mp 1)$, and
\begin{equation}
k^\pm=\frac{E_q\pm k^3}{\sqrt{2}},\;\;\; k_\perp^\mu=(0,k^1,k^2,0). \label{kpm}
\end{equation}
 The function $K(\vec{k})$ is a quantity proportional to the $B$ meson wave function
\begin{equation}
K(\vec{k})=\frac{2N_B\Psi_0(\vec{k})}{\sqrt{E_qE_Q(E_q+m_q)(E_Q+m_Q)}} \label{wave-k}
\end{equation}
and $\Psi_0(\vec{k})$ is the $B$ meson wave function in the rest frame with
\begin{equation}\label{psi0}
\Psi_0(\vec{k})=a_1 e^{a_2|\vec{k}|^2+a_3|\vec{k}|+a_4}
\end{equation}
where the parameters $a_i$ ($i=1,\cdots, 4$) are \cite{SY2017}
\begin{eqnarray}
	&&a_1=4.55_{-0.30}^{+0.40}\,\mathrm{GeV}^{-3/2},\quad\;
	a_2=-0.39_{-0.20}^{+0.15}\,\mathrm{GeV}^{-2},\nonumber\\
	&& a_3=-1.55\pm 0.20\,\mathrm{GeV}^{-1},\quad   a_4=-1.10_{-0.05}^{+0.10}.
\end{eqnarray}

The light-cone coordinate wave function for pion can be defined by \cite{bra1990,bal1999}
\begin{eqnarray}
\langle \pi(p_\pi)|\bar{q}(y)_\rho q'(0)_\delta |0\rangle &=&\int dx d^2k_{q\perp}e^{i(x p_\pi\cdot y-y_{\perp}\cdot k_{q\perp})}\nonumber\\
&&\times\Phi^\pi_{\delta\rho}
\end{eqnarray}
and $\Phi^\pi_{\delta\rho}$ is the spinor wave function
\begin{eqnarray} \label{pion-spinor1}
\Phi^\pi_{\delta\rho}&=&\frac{if_\pi}{4}\Bigg\{\not{p}_\pi\gamma_5\phi_\pi(x,k_{q\perp})
  -\mu_\pi\gamma_5\Bigg(\phi^\pi_P(x,k_{q\perp})\nonumber\\
   && -\sigma_{\mu\nu}p_\pi^\mu y^\nu\frac{\phi^\pi_\sigma(x,k_{q\perp})}{6}\Bigg)\Bigg\}_{\delta\rho}
\end{eqnarray}
with $f_\pi$ being the pion decay constant, $\mu_\pi$ the chiral parameter, and $\phi_\pi$, $\phi^\pi_P$, and $\phi^\pi_\sigma$ are twist-2 and twist-3 distribution functions, respectively. In the momentum space $\Phi^\pi_{\delta\rho}$ can be written as \cite{bf2001,wy2002}
\begin{eqnarray} \label{pion-spinor2}
\Phi^\pi_{\delta\rho}&=&\frac{if_\pi}{4}\Bigg\{\not{p}_\pi\gamma_5\phi_\pi(x,k_{q\perp})
  -\mu_\pi\gamma_5\Bigg(\phi^\pi_P(x,k_{q\perp})\nonumber\\
   && -i\sigma_{\mu\nu}\frac{p_\pi^\mu \bar{p}_\pi^\nu}{p_\pi\cdot \bar{p}_\pi} \frac{\phi'^\pi_\sigma(x,k_{q\perp})}{6}\nonumber\\
   &&+i  \sigma_{\mu\nu}p_\pi^\mu\frac{\phi^\pi_\sigma(x,k_{q\perp})}{6}\frac{\partial}{\partial k_{q\perp\nu}}\Bigg)\Bigg\}_{\delta\rho}
\end{eqnarray}
where $\bar{p}_\pi =(E_\pi,-\vec{p}_\pi)$ with $E_\pi$ and $\vec{p}_\pi$ being the energy and momentum of pion, respectively,  and $\phi'^{\pi}_\sigma (x,k_{q\perp})=\partial \phi^\pi_\sigma (x,k_{q\perp})/\partial x$.

The light-cone wave functions for the other light pseudoscalar mesons can be defined similarly as that of pion.

\subsection{ The scheme for $\eta-\eta' $ mixing}

The mixing scheme for $\eta$ and $\eta'$ mesons suggested by Feldmann, Kroll, and Stech in Refs. \cite{FKS1,FKS2} is considered in this work when calculating processes involving $\eta$ and $\eta'$ mesons. In this mixing scheme, the physical $\eta$ and $\eta'$ mesons are written as
\begin{equation}
	\left(\begin{array}{c}
		\eta \\
		\eta^{\prime}
	\end{array}\right)=U(\phi)\left(\begin{array}{c}
		\eta_q \\
		\eta_s
	\end{array}\right)=\left(\begin{array}{cc}
		\cos \phi & -\sin \phi \\
		\sin \phi & \cos \phi
	\end{array}\right)\left(\begin{array}{l}
		\eta_q \\
		\eta_s
	\end{array}\right)
\end{equation}
where $\eta_q=(u \bar{u}+d \bar{d}) / \sqrt{2}$, $\eta_s=s \bar{s}$, and $\phi$ is the mixing angle. We define the decay constants for the pseudoscalars involving $\eta$ and $\eta'$ as follows:
\begin{equation}
\langle 0|j^{q\mu}_5|\eta_q(p)\rangle =if_q p^\mu, ~~~\langle 0|j^{s\mu}_5|\eta_s(p)\rangle =if_s p^\mu
\end{equation}
where $j^{q\mu}_5=\frac{\bar{u}\gamma^\mu \gamma_5 u+\bar{d}\gamma^\mu \gamma_5 d } {\sqrt{2}}$ and
$j^{s\mu}_5=\bar{s}\gamma^\mu \gamma_5 s$,
\begin{equation}
\begin{array}{ll}
\langle 0|\bar{q}\gamma^\mu\gamma_5q|\eta(p)\rangle =if_\eta^q p^\mu, &\langle 0|j^{s\mu}_5|\eta(p)\rangle =if_\eta^s p^\mu, \\
\langle 0|\bar{q}\gamma^\mu\gamma_5q|\eta'(p)\rangle =if_{\eta'}^q p^\mu, &\langle 0|j^{s\mu}_5|\eta'(p)\rangle =if_{\eta'}^s p^\mu
\end{array}
\end{equation}
where $q=u$ and $d$. The relations between the decay constants hold
\begin{equation}
	\begin{array}{ll}
		f_\eta^q=\frac{f_q}{\sqrt{2}} \cos \phi, & f_\eta^s=-f_s \sin \phi, \\
		&  \\
		f_{\eta^{\prime}}^q=\frac{f_q}{\sqrt{2}} \sin \phi, & f_{\eta^{\prime}}^s=f_s \cos \phi
	\end{array}
\end{equation}
where isospin symmetry is taken into account. The values of these decay constants are \cite{FKS1,FKS2}
\begin{equation}
	\begin{gathered}
		f_q=(1.07 \pm 0.02) f_\pi, \quad f_s=(1.34 \pm 0.06) f_\pi , \\
		\phi=39.3^{\circ} \pm 1.0^{\circ}
	\end{gathered}
\end{equation}
where $f_\pi = 0.130 \;\mathrm{GeV}$. The chiral masses for $\eta_q$ and $\eta_s$ are $m_0^q$ and $m_0^s$, respectively, which replace $\mu_\pi$ in Eqs. (\ref{pion-spinor1}) and (\ref{pion-spinor2}) when considering the light-cone wave functions of $\eta_q$ and $\eta_s$. The values of them are $m_0^q = 1.07 \;\mathrm{GeV}$, $m_0^s = 1.82\; \mathrm{GeV}$, which can be obtained by \cite{FKS1,FKS2}

\begin{equation}
	\begin{aligned}
		& m_0^q=\frac{1}{2 m_q}\left(U_{11}-\frac{\sqrt{2} f_s}{f_q} U_{12}\right), \\
		& m_0^s=\frac{1}{2 m_s}\left(U_{22}-\frac{f_q}{\sqrt{2} f_s} U_{21}\right)
	\end{aligned}
\end{equation}
\vspace{1em}
\begin{equation}
	\begin{aligned}
		U_{11}= & m_\eta^2 \cos^2 \phi  +  m_{\eta^{\prime}}^2\sin^2 \phi , \\
		U_{12}=  U_{21}& = \left(m_{\eta^{\prime}}^2-m_\eta^2\right)\cos \phi \sin \phi \\
		U_{22}= & m_\eta^2 \sin^2 \phi  +  m_{\eta^{\prime}}^2\cos^2 \phi
	\end{aligned}
\end{equation}
\vspace{1em}

where $m_\eta= 0.548 \;\mathrm{GeV}, \quad m_{\eta'}=0.958 \;\mathrm{GeV}$. And the quarks mass are $m_q = 0.0056\mathrm{GeV},\quad m_s=0.137 \mathrm{GeV}$, which are consistent with the input parameters in the work of Ref. \cite{Ball-Braun2006}.


\subsection{The leading order contribution}

The diagrams for the hard amplitude at leading order (LO) in QCD are shown in Fig. \ref{fig1}. Transverse momenta of quarks and gluons are kept in the calculation, and double logarithms as $\alpha_s(\mu)\mbox{ln}^2 k_\text{T}/\mu$ appear in higher order radiative corrections in QCD, which can be resummed into the Sudakov factor \cite{liyu1996-1,liyu1996-2}.  The double logarithms as $\alpha_s(\mu)\mbox{ln}^2 x$ can be resummed into the threshold factor \cite{lihn2002}, where $x$ is the momentum fraction of gluons or quarks in the longitudinal direction.
Diagrams (a), (b), (g), and (h) in Fig. \ref{fig1} are factorizable diagrams and (c), (d), (e), and (f) are the  nonfactorizable ones. The amplitude contributed by (a) and (b) with the insertion of operators of the $(V-A)(V-A)$ current is

\begin{widetext}
\begin{eqnarray}\label{fe}
&F_e &=-i\frac{4\pi^2}{N_c^2}f_B f_{M_1} f_{M_2} m_B\int dk_{\perp}k_{\perp}\int_{x^d}^{x^u}dx\int_0^1 dx_1 \int_0^\infty bdbb_1db_1 (\frac{1}{2}m_B+\frac{|\vec{k}_{\perp}|^2}{2x^2m_B})K(\vec{k})(E_Q+m_Q)\nonumber\\
&&\times J_0(k_{\perp}b) \Bigg\{\alpha_s(\mu_{e}^1)\Bigg(2m_B[E_q(1+x_1)+k^3(1-x_1)] \phi_{M_1}(\bar{x}_1,b_1)+2\mu_{M_1}[E_q(1-2x_1)-k^3]\phi^P_{M_1}(\bar{x}_1,b_1)\nonumber\\
&&-\frac{1}{3}\mu_{M_1}[E_q(1-2 x_1)-k^3]\phi'^\sigma_{M_1}(\bar{x}_1,b_1)\Bigg) h_e^1(x,x_1,b,b_1)S_t(x_1)\exp[-S_{B}(\mu_{e}^1)-S_{M_1}(\mu_{e}^1)] \nonumber\\
&&+\alpha_s(\mu_{e}^2) [4\mu_{M_1}(E_q-k^3)]\phi^P_{M_1}(\bar{x}_1,b_1)h_e^2(x,x_1,b,b_1) S_t(x)\exp[-S_{B}(\mu_{e}^2)-S_{M_1}(\mu_{e}^2)]\Bigg\}
\end{eqnarray}
where $\bar{x}_i=1-x_i$, for $i=1,\; 2$. The integral with respect to the momentum fraction $x$ for the light quark in the $B$ meson along the light-cone direction is limited from $x^d$ to $x^u$, where $x^{u,d}=1/2\pm\sqrt{1/4-|\vec{k_\perp}|^2/m_B^2}$ \cite{Lu-Yang2021}.

\begin{figure}[bth]
\begin{center}
	\epsfig{file=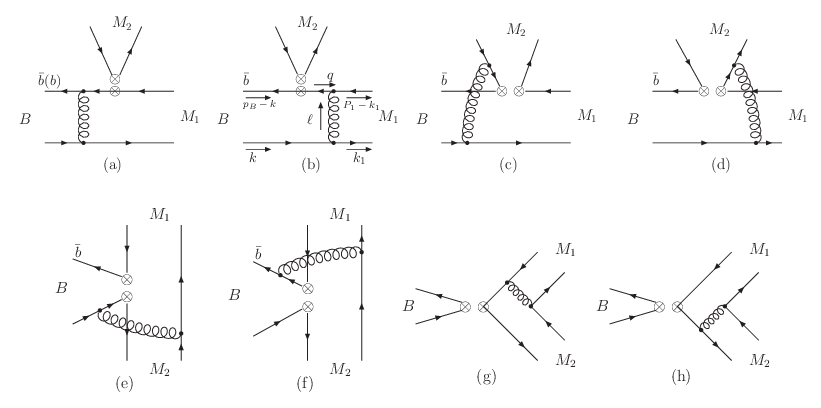,width=15cm,height=8cm}
  \caption{Diagrams contributing to the $B\rightarrow M_1 M_2$ decays at leading order in QCD, where the diagrams can be classified into four types: factorizable emission diagrams (a, b), nonfactorizable emission diagrams (c, d), factorizable annihilation diagrams (e, f), and nonfactorizable annihilation diagrams (g, h).} \label{fig1}
\end{center}
\end{figure}

The operators of the $(S+P)(S-P)$ current which comes from the Fierz transformation of the operators of the $(V-A)(V+A)$ current contribute as
\begin{eqnarray}\label{feP}
&F_e^P &=-i\frac{4\pi^2}{N_c^2}f_B f_{M_1} f_{M_2}  \mu_{M_2}\int dk_{\perp}k_{\perp}\int_{x^d}^{x^u}dx\int_0^1 dx_1 \int_0^\infty bdbb_1db_1 (\frac{1}{2}m_B+\frac{|\vec{k}_{\perp}|^2}{2x^2m_B})K(\vec{k})(E_Q+m_Q)\nonumber\\
&&\times J_0(k_{\perp}b) \Bigg\{\alpha_s(\mu_{e}^1)\Bigg(4m_B(E_q+k^3)\phi_{M_1}(\bar{x}_1,b_1)+4\mu_{M_1}[E_q(x_1+2)-k^3 x_1]\phi^P_{M_1}(\bar{x}_1,b_1)\nonumber\\
&&-\frac{2}{3}\mu_{M_1}[k^3(x_1-2)-E_q x_1]\phi'^\sigma_{M_1}(\bar{x}_1,b_1)\Bigg) h_e^1(x,x_1,b,b_1) S_t(x_1)\exp[-S_{B}(\mu_{e}^1)-S_{M_1}(\mu_{e}^1)] \nonumber\\
&&+\alpha_s(\mu_{e}^2) [ 8\mu_{M_1}(E_q-k^3) ]\phi^P_{M_1}(\bar{x}_1,b_1)h_e^2(x,x_1,b,b_1) S_t(x)\exp[-S_{B}(\mu_{e}^2)-S_{M_1}(\mu_{e}^2)]\Bigg\}.
\end{eqnarray}
The contributions of diagrams of Figs. \ref{fig1} (c) and \ref{fig1}(d) are
\begin{eqnarray}\label{Me}
&M_e &=-i\frac{4\pi^2}{N_c^2}f_B f_{M_1} f_{M_2} m_B\int dk_{\perp}k_{\perp}\int_{x^d}^{x^u}dx\int_0^1 dx_1 dx_2 \int_0^\infty bdbb_2db_2 (\frac{1}{2}m_B+\frac{|\vec{k}_{\perp}|^2}{2x^2m_B})K(\vec{k})(E_Q+m_Q)\nonumber\\
&&\times J_0(k_{\perp}b)\phi_{M_2}(\bar{x}_2,b_2) \Bigg\{\alpha_s(\mu_{d}^2)
\Bigg(-2m_B(x_2-1) (E_q+k^3)\phi_{M_1}(\bar{x}_1,b)-2\mu_{M_1} x_1(E_q-k^3)\phi^P_{M_1}(\bar{x}_1,b)\nonumber \\
&&-\frac{1}{3}\mu_{M_1} x_1(E_q-k^3)\phi'^\sigma_{M_1}(\bar{x}_1,b)\Bigg) h_d^1(x,x_1,x_2,b,b_2) S_t(x_1)\exp[-S_{B}(\mu_{d}^1)-S_{M_1}(\mu_{d}^1)-S_{M_2}(\mu_{d}^1)] \nonumber\\
&&+\alpha_s(\mu_{d}^2) \Bigg(-2m_B[E_q(x_1+x_2)+k^3(x_2-x_1) \phi_{M_1}(\bar{x}_1,b)+2\mu_{M_1} x_1(E_q+k^3)\phi^P_{M_1}(\bar{x}_1,b)\nonumber\\
&&-\frac{1}{3}\mu_{M_1} x_1(E_q+k^3)\phi'^\sigma_{M_1}(\bar{x}_1,b)\Bigg) h_d^2(x,x_1,x_2,b,b_2)\times S_t(x_1)\exp[-S_{B}(\mu_{d}^2)-S_{M_1}(\mu_{d}^2)-S_{M_2}(\mu_{d}^2)]\Bigg\}
\end{eqnarray}
for the operators of the $(V-A)(V-A)$ current, and
\begin{eqnarray}\label{MeP}
&M_e^P &=-i\frac{4\pi^2}{N_c^2}f_B f_{M_1} f_{M_2} m_B\int dk_{\perp}k_{\perp}\int_{x^d}^{x^u}dx\int_0^1 dx_1 dx_2 \int_0^\infty bdbb_2db_2 (\frac{1}{2}m_B+\frac{|\vec{k}_{\perp}|^2}{2x^2m_B})K(\vec{k})(E_Q+m_Q)\nonumber\\
&&\times J_0(k_{\perp}b)\phi_{M_2}(\bar{x}_2,b_2) \Bigg\{\alpha_s(\mu_{d}^2)
\Bigg(-2m_B(E_q(x_1-x_2+1)-k^3(x_1+x_2-1))\phi_{M_1}(\bar{x}_1,b)\nonumber\\
&&+2\mu_{M_1} x_1(E_q+k^3)\phi^P_{M_1}(\bar{x}_1,b)-\frac{1}{3}\mu_{M_1} x_1(E_q+k^3) \phi'^\sigma_{M_1}(\bar{x}_1,b)\Bigg) h_d^1(x,x_1,x_2,b,b_2)\nonumber\\
&&\times S_t(x_1)\exp[-S_{B}(\mu_{d}^1)-S_{M_1}(\mu_{d}^1)-S_{M_2}(\mu_{d}^1)] +\alpha_s(\mu_{d}^2) \Bigg(2m_Bx_2(E_q+k^3)\phi_{M_1}(\bar{x}_1,b)\nonumber
\end{eqnarray}	
\begin{eqnarray}
&&-2\mu_{M_1} x_1(E_q-k^3)\phi^P_{M_1}(\bar{x}_1,b)-\frac{1}{3}\mu_{M_1} x_1(E_q-k^3)\phi'^\sigma_{M_1}(\bar{x}_1,b)\Bigg) h_d^2(x,x_1,x_2,b,b_2)\nonumber\\
&&\times S_t(x_1)\exp[-S_{B}(\mu_{d}^2)-S_{M_1}(\mu_{d}^2)-S_{M_2}(\mu_{d}^2)]\Bigg\}
\end{eqnarray}
for the insertion of Fierz transformed operators of the $(S+P)(S-P)$ current. The following is
for the contribution of the operators of the $(V-A)(V+A)$ current
\begin{eqnarray}
&M_e^R &=-i\frac{4\pi^2}{N_c^2}f_Bf_{M_1} f_{M_2} \int dk_{\perp}k_{\perp}\int_{x^d}^{x^u}dx\int_0^1 dx_1 dx_2 \int_0^\infty bdbb_2db_2 (\frac{1}{2}m_B+\frac{|\vec{k}_{\perp}|^2}{2x^2m_B})K(\vec{k})(E_Q+m_Q)\nonumber\\
&&\times J_0(k_{\perp}b) \Bigg\{\alpha_s(\mu_{d}^2)
\Bigg[\frac{1}{3}\mu_{M_2} m_B(E_q+k^3)\phi_\pi(\bar{x}_1,b)\Big((x_2-1)\phi'^\sigma_{M_2}(\bar{x}_2,b_2) -6(x_2-1)\phi^P_{M_2}(\bar{x}_2,b_2)\Big) \nonumber\\
&& +\frac{1}{3}\mu_{M_1} \mu_{M_2}\phi^P_{M_1}(\bar{x}_1,b)\Big([E_q(x_1+x_2-1)+k^3(-x_1 +x_2-1)]\phi'^\sigma_{M_2}(\bar{x}_2,b_2)+6[E_q(x_1-x_2+1)-k^3(x_1\nonumber\\
&&+x_2-1)]\phi^P_{M_2}(\bar{x}_2,b_2)\Big) -\frac{1}{18}\mu_{M_1}\mu_{M_2}\phi'^\pi_\sigma(\bar{x}_1,b) \Big( [E_q(x_1-x_2+1)-k^3(x_1+x_2-1)]\nonumber\\
&&\cdot \phi'^\sigma_{M_2}(\bar{x}_2,b_2) +6[E_q(x_1+x_2-1)+k^3(-x_1+x_2-1)]\phi^P_{M_2}(\bar{x}_2,b_2) \Big)\Bigg] h_d^1(x,x_1,x_2,b,b_2)\nonumber\\
&&\times S_t(x_1)\exp[-S_{B}(\mu_{d}^1)-S_{M_1}(\mu_{d}^1)-S_{M_2}(\mu_{d}^1)] +\alpha_s(\mu_{d}^2) \Bigg[-\frac{1}{3}\mu_{M_2}m_B(E_q+k^3)\Big(x_2\phi'^\sigma_{M_2}(\bar{x}_2,b_2)\nonumber\\
&& +6x_2\phi^P_{M_2}(\bar{x}_2,b_2)\Big)\phi_{M_1}(\bar{x}_1,b)+\frac{1}{3}\mu_{M_1} \mu_{M_2}\phi^P_{M_1}(\bar{x}_1,b)  \Big([E_q(x_1-x_2) -k^3(x_1+x_2)]\phi'^\sigma_{M_2}(\bar{x}_2,b_2)\nonumber\\
&& +6[E_q(x_1+x_2)+k^3(x_2-x_1)]\phi^P_{M_2}(\bar{x}_2,b_2)\Big) -\frac{1}{18}\mu_{M_1} \mu_{M_2}\phi'^\sigma_{M_1}(\bar{x}_1,b)\Big( [E_q(x_1+x_2)+k^3(x_2-x_1)]\nonumber\\
&&\cdot \phi'^\sigma_{M_2}(\bar{x}_2,b_2) +6[E_q(x_2-x_1)+k^3(x_1+x_2)]  \phi^P_{M_2}(\bar{x}_2,b_2) \Big)\Bigg]h_d^2(x,x_1,x_2,b,b_2)S_t(x_1)\nonumber\\
&&\times\exp[-S_{B}(\mu_{d}^2)-S_{M_1}(\mu_{d}^2)-S_{M_2}(\mu_{d}^2)]\Bigg\}.
\end{eqnarray}

The contributions of Figs. \ref{fig1} (e) and \ref{fig1} (f) are
\begin{eqnarray}\label{Ma}
M_a &=&-i\frac{4\pi^2}{N_c^2 }f_B f_{M_1} f_{M_2} \int dk_{\perp}k_{\perp}\int_{x^d}^{x^u}dx\int_0^1 dx_1 dx_2 \int_0^\infty bdbb_1db_1 (\frac{1}{2}m_B+\frac{|\vec{k}_{\perp}|^2}{2x^2m_B})K(\vec{k}) (E_Q+m_Q)\nonumber\\
&&\times J_0(k_{\perp}b)\Bigg\{\alpha_s(\mu_{f}^2)
\Bigg[-2m_B^2(x_2-1)(E_q+k^3)\phi_{M_1}(\bar{x}_1,b_1)\phi_{M_2}(\bar{x}_2,b_1)+\frac{1}{3}\mu_{M_1}\mu_{M_2}\phi^P_{M_1}(\bar{x}_1,b_1)\nonumber\\
&&\cdot\Big([-E_q(x_1+x_2-1)+k^3(x_1-x_2+1)]\phi'^\sigma_{M_2}(\bar{x}_2,b_1) +6[E_q(x_1-x_2+1)-k^3(x_1+x_2-1)]\phi^P_{M_2}(\bar{x}_2,b_1)\Big)\nonumber\\
&& -\frac{1}{18}\mu_{M_1}\mu_{M_2}\phi'^\sigma_{M_1}(\bar{x}_1,b_1)\Big([E_q(x_1-x_2+1)-k^3(x_1 +x_2-1)]\phi'^\sigma_{M_2}(\bar{x}_2,b_1)-6[E_q(x_1+x_2-1)+k^3(-x_1\nonumber\\
&&+x_2-1)]\phi^P_{M_2}(\bar{x}_2,b_1) \Big)\Bigg]h_f^1(x,x_1,x_2,b,b_1) S_t(x_1)S_t(x_2)\exp[-S_{B}(\mu_{f}^1)-S_{M_1}(\mu_{f}^1)-S_{M_2}(\mu_{f}^1)]
\nonumber \\
&&+\alpha_s(\mu_{d}^2) \Bigg[-2m_B^2x_1(E_q-k^3)\phi_{M_1}(\bar{x}_1,b_1)\phi_{M_2}(\bar{x}_2,b_1) -\frac{1}{3}\mu_{M_1}\mu_{M_2}\phi^P_{M_1}(\bar{x}_1,b_1) \Big([E_q(x_1+x_2-1)+k^3(-x_1\nonumber
\end{eqnarray}	
\begin{eqnarray}
&&+x_2+1)]\phi'^\sigma_{M_2}(\bar{x}_2,b_1)+6[E_q(x_1-x_2+3)-k^3(x_1+x_2-1)]\phi^P_{M_2}(\bar{x}_2,b_1)\Big) +\frac{1}{18}\mu_{M_1}\mu_{M_2}\phi'^\sigma_{M_1}(\bar{x}_1,b_1) \nonumber\\
&&\cdot\Big([E_q(x_1-x_2-1)-k^3(x_1+x_2-1)]\phi'^\sigma_{M_2}(\bar{x}_2,b_1) +6[E_q(x_1+x_2-1)+k^3(-x_1+x_2-3)]\phi^P_{M_2}(\bar{x}_2,b_1) \Big)\Bigg]\nonumber\\
&& \times h_f^2(x,x_1,x_2,b,b_1)\exp[-S_{B}(\mu_{f}^2)-S_{M_1}(\mu_{f}^2)-S_{M_2}(\mu_{f}^2)]\Bigg\}
\end{eqnarray}
for the $(V-A)(V-A)$ current, and
\begin{eqnarray}\label{MaP}
&M_a^P &=-i\frac{4\pi^2}{N_c^2 }f_B f_{M_1} f_{M_2} \int dk_{\perp}k_{\perp}\int_{x^d}^{x^u}dx\int_0^1 dx_1 dx_2 \int_0^\infty bdbb_1db_1 (\frac{1}{2}m_B+\frac{|\vec{k}_{\perp}|^2}{2x^2m_B})K(\vec{k})(E_Q+m_Q)\nonumber\\
&&\times J_0(k_{\perp}b) \Bigg\{\alpha_s(\mu_{f}^2)\Bigg[2m_B^2x_1(E_q-k^3)\phi_{M_1}(\bar{x}_1,b_1)\phi_{M_2}(\bar{x}_2,b_1)+\frac{1}{3}\mu_{M_1}\mu_{M_2}\phi^P_{M_1}(\bar{x}_1,b_1)\Big([E_q(x_1+x_2-1)\nonumber\\
&&+k^3(-x_1+x_2-1)]\phi'^\sigma_{M_2}(\bar{x}_2,b_1)+6[E_q(x_1-x_2+1)-k^3(x_1+x_2-1)]\phi^P_{M_2}(\bar{x}_2,b_1)\Big) \nonumber\\
&& -\frac{1}{18}\mu_{M_1}\mu_{M_2}\phi'^\sigma_{M_1}(\bar{x}_1,b_1)\Big([E_q(x_1-x_2+1) -k^3(x_1+x_2-1)]\phi'^\sigma_{M_2}(\bar{x}_2,b_1)  +6[E_q(x_1+x_2-1)\nonumber\\
&&+k^3(-x_1+x_2-1)]\phi^P_{M_2}(\bar{x}_2,b_1) \Big)\Bigg] h_f^1(x,x_1,x_2,b,b_1)S_t(x_1)S_t(x_2)\exp[-S_{B}(\mu_{f}^1)-S_{M_1}(\mu_{f}^1)-S_{M_2}(\mu_{f}^1)]\nonumber\\
&& +\alpha_s(\mu_{d}^2) \Bigg[2m_B^2(x_2-1)(E_q+k^3)\phi_{M_1}(\bar{x}_1,b_1)\phi_{M_2}(\bar{x}_2,b_1) +\frac{1}{3}\mu_{M_1}\mu_{M_2}\phi^P_{M_1}(\bar{x}_1,b_1) \nonumber\\
&&\Big( [E_q(x_1+x_2-1)+k^3(-x_1+x_2-3)]\phi'^\sigma_{M_2}(x_2,b_2-6[E_q(x_1-x_2+3)-k^3(x_1+x_2-1)]\phi^P_{M_2}(\bar{x}_2,b_1)\Big)\nonumber\\
&& +\frac{1}{18}\mu_{M_1}\mu_{M_2}\phi'^\sigma_{M_1}(\bar{x}_1,b_1) \Big([E_q(x_1-x_2-1)-k^3(x_1+x_2-1)]\phi'^\sigma_{M_2}(\bar{x}_2,b_1) -6[E_q(x_1+x_2-1)+k^3(-x_1+x_2\nonumber\\
&&+1)]\phi^P_{M_2}(\bar{x}_2,b_1) \Big)\Bigg] h_f^2(x,x_1,x_2,b,b_1)\exp[-S_{B}(\mu_{f}^2)-S_{M_1}(\mu_{f}^2)-S_{M_2}(\mu_{f}^2)]\Bigg\}
\end{eqnarray}
for the $(S+P)(S-P)$ current,
\begin{eqnarray}\label{MaR}
&M_a^R &=-i\frac{4\pi^2}{N_c^2}f_B f_{M_1} f_{M_2} m_B\int dk_{\perp}k_{\perp}\int_{x^d}^{x^u}dx\int_0^1 dx_1 dx_2 \int_0^\infty bdbb_2db_2 (\frac{1}{2}m_B+\frac{|\vec{k}_{\perp}|^2}{2x^2m_B})K(\vec{k})(E_Q+m_Q)\nonumber\\
&&\times J_0(k_{\perp}b)\phi_{M_2}(\bar{x}_2,b_1) \Bigg\{\alpha_s(\mu_{f}^2)\Bigg(\frac{1}{3}\mu_{M_2}(E_q-k^3)\phi_{M_1}(\bar{x}_1,b_1)(x_2-1)\Big(\phi'^\sigma_{M_2}(\bar{x}_2,b_1) +6\phi^P_{M_2}(\bar{x}_2,b_1) \Big)\nonumber\\
&&-2 x_1\mu_{M_1}(E_q+k^3)\phi_{M_2}(\bar{x}_2,b_1)\phi^P_{M_1}(\bar{x}_1,b_1)+\frac{1}{3} x_1\mu_{M_1}(E_q+k^3)\phi_{M_2}(\bar{x}_2,b_1)\phi'^\sigma_{M_1}(\bar{x}_1,b_1)\Bigg)\nonumber\\
&&\times h_f^1(x,x_1,x_2,b,b_2)S_t(x_1)S_t(x_2)\exp[-S_{B}(\mu_{f}^1)-S_{M_1}(\mu_{f}^1)-S_{M_2}(\mu_{f}^1)] +\alpha_s(\mu_{f}^2) \Bigg(-\frac{1}{3}\mu_{M_2}[E_q(x_2+1)\nonumber\\
&&+k^3(x_2-1)]\Big(\phi'^\sigma_{M_2}(\bar{x}_2,b_1) +6\phi^P_{M_2}(\bar{x}_2,b_1)\Big)\phi_{M_1}(\bar{x}_1,b_1)-2 \mu_{M_1}[E_q(x_2-2)-k^3 x_2] \nonumber\\
&&\phi_{M_2}(\bar{x}_2,b_1)\phi^P_{M_1}(\bar{x}_1,b_1)+\frac{1}{3}\mu_{M_1} [E_q(x_2-2)-k^3x_2]\phi_{M_2}(\bar{x}_2,b_1)\phi'^\sigma_{M_1}(\bar{x}_1,b_1)\Bigg) h_f^2(x,x_1,x_2,b,b_2)\nonumber\\
&& \times\exp[-S_{B}(\mu_{f}^2)-S_{M_1}(\mu_{f}^2)-S_{M_2}(\mu_{f}^2)]\Bigg\}
\end{eqnarray}
for the $(V-A)(V+A)$ current.

For diagrams (g) and (h) in Fig. \ref{fig1}, the contributions of the operators of $(V-A)(V-A)$ always cancel each other if the wave functions of the light mesons are symmetric with respect to the momentum fractions $x_1$ and $x_2$. If it is not the case, there will be small residual contributions. The contributions of diagrams (g) and (h) with the operators $(V-A)(V-A)$ are
\begin{eqnarray}\label{fa}
&F_a &=-i\frac{8\pi  }{N_c^2}f_B f_{M_1} f_{M_2}  \int_0^1 dx_1 dx_2 \int_0^\infty b_1db_1b_2db_2
\Bigg\{\alpha_s(\mu_{a}^1)\Bigg(-m_B^2 (x_2-1) \phi_{M_1}(\bar{x}_1,b_1)\phi_{M_2}(\bar{x}_2,b_2)\nonumber\\
&&-\frac{1}{3} \mu_{M_1}\mu_{M_2}\Bigg[ x_2\phi'^\sigma_{M_2}(\bar{x}_2,b_2)+6(x_2-2)\phi^P_{M_2}(\bar{x}_2,b_2)\Bigg] \phi^P_{M_1}(\bar{x}_1,b_1)\Bigg)h_a^1(x_1,x_2,b_1,b_2)S_t(x_2) \exp[-S_{M_1}(\mu_{a}^1)\nonumber\\
&&-S_{M_2}(\mu_{a}^1)] +\alpha_s(\mu_{a}^2) \Bigg(-m_B^2 x_1\phi_{M_1}(\bar{x}_1,b_1)\phi_{M_2}(\bar{x}_2,b_2)   -2\mu_{M_1}\mu_{M_2}(x_2+1) \phi^P_{M_1}(\bar{x}_1,b_1)\phi^P_{M_2}(\bar{x}_2,b_2)\nonumber\\
&& +\frac{1}{3}\mu_{M_1}\mu_{M_2}  (x_1-1)\phi'^\sigma_{M_1}(\bar{x}_1,b_1)\phi^P_{M_2}(\bar{x}_2,b_2)\Bigg) h_a^2(x_1,x_2,b_1,b_2)S_t(x_1)\exp[-S_{M_1}(\mu_{a}^2)-S_{M_2}(\mu_{a}^2)]\Bigg\}.
\end{eqnarray}

The main contributions come from the operators of the $(S+P)(S-P)$ currents. The result is
\begin{eqnarray}\label{faP}
&F_a^P &=-i\frac{8\pi  }{N_c^2} \chi _B f_{M_1} f_{M_2}  \int_0^1 dx_1 dx_2 \int_0^\infty b_1db_1b_2db_2 \Bigg\{\alpha_s(\mu_{a}^1)\Bigg(-4\mu_{M_1}\phi^P_{M_1}(\bar{x}_1,b_1)\phi_{M_2}(\bar{x}_2,b_2)\nonumber\\
&&-\frac{1}{3} \mu_{M_2}\Bigg[ (x_2-1)\phi'^\sigma_{M_2}(\bar{x}_2,b_2)-6(x_2-1)\phi^P_{M_2}(\bar{x}_2,b_2)\Bigg] \phi_{M_1}(\bar{x}_1,b_1)\Bigg)h_a^1(x_1,x_2,b_1,b_2)S_t(x_2) \exp[-S_{M_1}(\mu_{a}^1)\nonumber\\
&&-S_{M_2}(\mu_{a}^1)] +\alpha_s(\mu_{a}^2) \Bigg(-4\mu_{M_2} \phi_{M_1}(\bar{x}_1,b_1)\phi^P_{M_2}(\bar{x}_2,b_2)-\frac{1}{3}\mu_{M_1} \Bigg[ x_1\phi'^\sigma_{M_1}(\bar{x}_1,b_1)+6x_1\phi^P_{M_1}(\bar{x}_1,b_1)\Bigg]\phi_{M_2}(\bar{x}_2,b_2)\Bigg)\nonumber \\
&& \times h_a^2(x_1,x_2,b_1,b_2)S_t(x_1)\exp[-S_{M_1}(\mu_{a}^2)-S_{M_2}(\mu_{a}^2)]\Bigg\},
\end{eqnarray}

where
\begin{eqnarray} \label{chiB}
	\chi _B &&= \pi f_B m_B \int dk_{\perp}k_{\perp}\int_{x^d}^{x^u}dx(\frac{1}{2}m_B+\frac{|\vec{k}_{\perp}|^2}{2x^2m_B})    K(\vec{k}) \Bigg[(E_q+m_q)(E_Q+m_Q)+(E_q^2-m_q^2)\Bigg].
\end{eqnarray}

\end{widetext}
In Eqs. (\ref{fe})$\--$(\ref{faP}), the Sudakov factors $\exp[-S_{B}(\mu)]$, $\exp[-S_{M_1}(\mu)]$ and $\exp[-S_{M_2}(\mu)]$ are associated with each meson at the relevant energy scale, which are given in the Appendix \ref{a}. $\phi_M(x,b)$, $\phi^P_M(x,b)$, and $\phi^\sigma_M(x,b)$ are the wave functions of light meson in $b$ space, with $\vec{b}$ being the conjugate variable of the transverse momentum $\vec{k}_{\perp}$, which can be found in Appendix \ref{b}. The functions $h_i$'s are Fourier transformations of the hard amplitudes, which are
\begin{widetext}
	\begin{eqnarray}
		h_e^1(x,x_1,b,b_1)&=&K_0(\sqrt{xx_1}m_B b)\Big[\theta (b-b_1)I_0(\sqrt{x_1}m_B b_1)K_0(\sqrt{x_1}m_B b)
		+\theta(b_1-b) I_0(\sqrt{x_1}m_B b)K_0(\sqrt{x_1}m_B b_1)\Big],\nonumber\\
		&&
	\end{eqnarray}
	\begin{eqnarray}
		h_e^2(x,x_1,b,b_1)&=&K_0(\sqrt{xx_1}m_B b)  \Big[\theta(b-b_1)I_0(\sqrt{x}m_B b_1)K_0(\sqrt{x}m_B b)+\theta(b_1-b)  I_0(\sqrt{x}m_B b)K_0(\sqrt{x}m_B b_1)\Big],\nonumber\\
		&&
	\end{eqnarray}
	\begin{eqnarray}
		h_d^1(x,x_1,x_2,b,b_2)&=&K_0(-i\sqrt{x_1(1-x_2)}m_Bb_2)\Big[\theta(b_2-b)I_0(\sqrt{xx_1}m_B b)K_0(\sqrt{xx_1}m_Bb_2)+\theta(b-b_2) I_0(\sqrt{xx_1}m_Bb_2)\nonumber\\
		&& \times K_0(\sqrt{xx_1}m_B b)\Big],
	\end{eqnarray}
	\begin{eqnarray}
		h_d^2(x,x_1,x_2,b,b_2)&=&K_0(-i\sqrt{x_1x_2}m_B b_2)\Big[\theta(b_2-b)I_0(\sqrt{xx_1}m_B b)K_0(\sqrt{xx_1}m_B b_2)+\theta(b-b_2)I_0(\sqrt{xx_1}m_B b_2)\nonumber\\
		&&\times K_0(\sqrt{xx_1}m_B b)\Big],
	\end{eqnarray}
	\begin{eqnarray}
		h_f^1(x_1,x_2,b,b_1)&=&K_0(-i\sqrt{x_1(1-x_2)}m_B b)\Big[\theta(b-b_1)I_0(-i\sqrt{x_1(1-x_2)}m_B b_1)K_0(-i\sqrt{x_1(1-x_2)}m_B b)\nonumber\\
		&&+\theta(b_1-b)I_0(-i\sqrt{x_1(1-x_2)}m_B b)K_0(-i\sqrt{x_1(1-x_2)}m_B b_1)\Big],
	\end{eqnarray}
	\begin{eqnarray}
		h_f^2(x_1,x_2,b,b_1)&=&K_0(\sqrt{1-x_2+x_1x_2}m_Bb)\Big[\theta(b-b_1)I_0(-i\sqrt{x_1(1-x_2)}m_B b_1)K_0(-i\sqrt{x_1(1-x_2)}m_B b)\nonumber\\
		&&+\theta(b_1-b)I_0(-i\sqrt{x_1(1-x_2)}m_B b)K_0(-i\sqrt{x_1(1-x_2)}m_B b_1)\Big],
	\end{eqnarray}
	\begin{eqnarray}
		h_a^1(x_1,x_2,b_1,b_2)&=&K_0(-i\sqrt{x_1(1-x_2)}m_B b_1)\Big[\theta(b_2-b_1)I_0(-i\sqrt{1-x_2}m_B b_1)K_0(-i\sqrt{1-x_2}m_B b_2)\nonumber\\
		&&+\theta(b_1-b_2)I_0(-i\sqrt{1-x_2}m_Bb_2)K_0(-i\sqrt{1-x_2}m_B b_1)\Big],
	\end{eqnarray}
	\begin{eqnarray}
		h_a^2(x_1,x_2,b_1,b_2)&=&K_0(-i\sqrt{x_1(1-x_2)} b_1)\Big[\theta(b_2-b_1)I_0(-i\sqrt{x_1}m_B b_1)K_0(-i\sqrt{x_1}m_B b_2)+\theta(b_1-b_2)\nonumber\\
		&&\cdot I_0(-i\sqrt{x_1}m_B b_2)K_0(-i\sqrt{x_1}m_B b_1)\Big].
	\end{eqnarray}
\end{widetext}

The hard scales for the amplitudes relevant to the diagrams in Fig. \ref{fig1} are taken as the largest mass scales involved in each diagram which help to suppress the largest logarithmic terms in the higher order corrections. They are
\begin{eqnarray}
	\mu_e^1 &=& \max(\sqrt{x_1}m_B,\sqrt{xx_1}m_B,1/b,1/b_1),\nonumber\\
	\mu_e^2 &=& \max(\sqrt{x}m_B,\sqrt{xx_1}m_B,1/b,1/b_1),\nonumber\\
	\mu_d^1 &=& \max(\sqrt{xx_1}m_B,\sqrt{x_1(1-x_2)}m_B,1/b_1,1/b_2),\nonumber\\
	\mu_d^2 &=& \max(\sqrt{xx_1}m_B,\sqrt{x_1x_2}m_B ,1/b_1,1/b_2),\nonumber\\
	\mu_f^1 &=& \max(\sqrt{x_1(1-x_2)}m_B,1/b_1,1/b_2),\nonumber
\end{eqnarray}
\begin{eqnarray}
	\mu_f^2 &=& \max(\sqrt{x_1(1-x_2)}m_B,\sqrt{1-x_2+x_1x_2}m_B,\nonumber\\
	     && \quad \quad  1/b_1,1/b_2),\nonumber\\
	\mu_a^1 &=& \max(\sqrt{1-x_2}m_B,\sqrt{x_1(1-x_2)}m_B,1/b_1,1/b_2),\nonumber\\
	\mu_a^2 &=& \max(\sqrt{x_1}m_B,\sqrt{x_1(1-x_2)}m_B,1/b_1,1/b_2).
\end{eqnarray}

The decay amplitudes of the $B\to M_1 M_2$ process can be expressed in terms of the matrix elements calculated based on the diagrams shown in Fig. \ref{fig1}, namely, Eqs. (\ref{fe})$\--$(\ref{faP}). The results are

\begin{widetext}
\begin{eqnarray}\label{Mp+p-}
\mathcal{M}\bar{(B^0}&&\rightarrow\pi^+ \pi^-)=F_{e,\pi \pi}\Big[ \xi_{ud}(\frac{1}{3}C_1+C_2)-\xi_{td}(\frac{1}{3}C_3+C_4+\frac{1}{3}C_9+C_{10}) \Big] -F^P_{e,\pi \pi} \xi_{td}\Big[ \frac{1}{3}C_5+C_6+\frac{1}{3}C_7+C_{8} \Big]\nonumber\\
&&+M_{e,\pi \pi}\Big[ \xi_{ud}(\frac{1}{3}C_1)-\xi_{td}(\frac{1}{3}C_3+\frac{1}{3}C_9) \Big]+M_{a,\pi \pi}\Big[ \xi_{ud}(\frac{1}{3}C_2)-\xi_{td}(\frac{1}{3}C_3  +\frac{2}{3}C_4-\frac{1}{6}C_9+\frac{1}{6}C_{10}) \Big] \nonumber\\
&&+M^R_{a,\pi \pi}\Big[ -\xi_{td}(\frac{1}{3}C_5-\frac{1}{6}C_7) \Big]+M^P_{a,\pi \pi}\Big[ -\xi_{td}(\frac{2}{3}C_6+\frac{1}{6}C_{8}) \Big]+F^P_{a,\pi \pi}\Big[
-\xi_{td}(\frac{1}{3}C_5+C_6-\frac{1}{6}C_7-\frac{1}{2}C_8) \Big],
\end{eqnarray}

\begin{eqnarray}\label{Mp0p0}
\sqrt{2}\mathcal{M}(\bar{B^0}&&\rightarrow\pi^0 \pi^0)=F_{e,\pi \pi}\Big[ -\xi_{ud}(C_1+\frac{1}{3}C_2)-\xi_{td}(\frac{1}{3}C_3+C_4+\frac{3}{2}C_7+\frac{1}{2}C_{8}-\frac{5}{3}C_9 -C_{10}) \Big]-F^P_{e,\pi \pi} \xi_{td}\Big[ \frac{1}{3}C_5+C_6 \nonumber\\
&&-\frac{1}{6}C_7-\frac{1}{2}C_{8} \Big]+M_{e,\pi \pi}\Big[ -\xi_{ud}(\frac{1}{3}C_2)-\xi_{td}(\frac{1}{3}C_3-\frac{1}{6}C_9-\frac{1}{2}C_{10}) \Big] +M^P_{e,\pi \pi}\Big[ -\xi_{td}(-\frac{1}{2}C_8) \Big] \nonumber\\
&&+M_{a,\pi \pi}\Big[ \xi_{ud}(\frac{1}{3}C_2)-\xi_{td}(\frac{1}{3}C_3+\frac{2}{3}C_4-\frac{1}{6}C_9  +\frac{1}{6}C_{10}) \Big]+M^R_{a,\pi \pi}\Big[ -\xi_{td}(\frac{1}{3}C_5-\frac{1}{6}C_7) \Big]\nonumber\\
&&+M^P_{a,\pi \pi}\Big[ -\xi_{td}(\frac{2}{3}C_6+\frac{1}{6}C_{8}) \Big] +F^P_{a,\pi \pi}\Big[ -\xi_{td}(\frac{1}{3}C_5+C_6-\frac{1}{6}C_7-\frac{1}{2}C_8) \Big],
\end{eqnarray}
and
\begin{eqnarray}\label{Mp-p0}
\sqrt{2}\mathcal{M}(B^-&&\rightarrow\pi^- \pi^0)=F_{e,\pi \pi}\Big[ \xi_{ud}(\frac{4}{3}C_1+\frac{4}{3}C_2)-\xi_{td}(2C_9-\frac{3}{2}C_7-\frac{1}{2}C_8+2C_{10}) \Big] -F^P_{e,\pi \pi} \xi_{td}\Big[ \frac{1}{2}C_7+\frac{3}{2}C_8 \Big]\nonumber\\
&&+M_{e,\pi \pi}\Big[ \xi_{ud}(\frac{1}{3}C_1+\frac{1}{3}C_2)-\xi_{td}(\frac{1}{2}C_9+\frac{1}{2}C_{10}) \Big]+M^P_{e,\pi \pi}\Big[ -\xi_{td}(\frac{1}{2}C_{8}) \Big]
\end{eqnarray}

\begin{eqnarray}\label{MK0p-}
\mathcal{M} (B^-&&\rightarrow\bar{K^0} \pi^-)=\xi_{us}\Big[ (\frac{1}{3}C_1+C_2)F_{a,\pi K}+ \frac{1}{3}C_1M_{a,\pi K}  \Big] -\xi_{ts}\Big[ (\frac{1}{3}C_3+C_4-\frac{1}{6}C_9-\frac{1}{2}C_{10}) F_{e,\pi K}  \nonumber\\
&& + (\frac{1}{3}C_5+C_6-\frac{1}{6}C_7-\frac{1}{2}C_{8})F^P_{e,\pi K}  +(\frac{1}{3}C_3-\frac{1}{6}C_9) M_{e,\pi K} + (\frac{1}{3}C_5 -\frac{1}{6}C_7)M^R_{e,\pi K} \nonumber\\
&&+(\frac{1}{3}C_3+C_4+\frac{1}{3}C_9+C_{10})F_{a,K \pi} +(\frac{1}{3}C_5+C_6 +\frac{1}{3}C_7+C_{8})F^P_{a,K \pi}+(\frac{1}{3}C_3+\frac{1}{3}C_9)M_{a,K \pi}+(\frac{1}{3}C_5 +\frac{1}{3}C_7)M^R_{a,K \pi} \Big]  \nonumber\\
\end{eqnarray}

\begin{eqnarray}\label{MK-p0}
\sqrt{2}\mathcal{M} (B^-&&\rightarrow K^- \pi^0)=\xi_{us}\Big[ (C_1+\frac{1}{3}C_2)F_{e,K \pi}+(\frac{1}{3}C_1+C_2)F_{e,\pi K} + (\frac{1}{3}C_2)M_{e,K \pi} + (\frac{1}{3}C_1) M_{e, \pi K} + (C_1+\frac{1}{3}C_2)F_{a,K \pi}\nonumber\\
&&  + (\frac{1}{3}C_1) M_{a, \pi K} \Big]   -\xi_{ts}\Big[  F_{e,\pi K}(\frac{1}{3}C_3+C_4+\frac{1}{3}C_9+C_{10}) + \frac{3}{2}F_{e,K \pi}( -C_7 -\frac{1}{3}C_8 + C_9 + \frac{1}{3}C_{10})   \nonumber\\
&& +F^P_{e,\pi K}(\frac{1}{3}C_5+C_6+\frac{1}{3}C_7+ C_{8})  +\frac{1}{2}C_{10}M_{e,K \pi}+\frac{1}{2}C_8M^P_{e,K \pi} +(\frac{1}{3}C_3+\frac{1}{3}C_9)M_{e,\pi K}\nonumber\\
&&+(\frac{1}{3}C_5+\frac{1}{3}C_7)M^R_{e,\pi K} + (\frac{1}{3}C_3+C_4+\frac{1}{3}C_9+C_{10})F_{a,K \pi}+(\frac{1}{3}C_5+C_6+\frac{1}{3}C_7+C_{8})F^P_{a,K \pi} \nonumber\\
&&+(\frac{1}{3}C_3+\frac{1}{3}C_9)M_{a,K \pi}+(\frac{1}{3}C_5+\frac{1}{3}C_7)M^R_{a,K \pi} \Big],\nonumber\\
&&
\end{eqnarray}

\begin{eqnarray}\label{MK-p+}
\mathcal{M} (\bar{B^0}&&\rightarrow K^- \pi^+)=\xi_{us}\Big[ (\frac{1}{3}C_1+C_2)F_{e,\pi K}+ \frac{1}{3} C_1 M_{e,\pi K}\Big] -\xi_{ts}\Big[ (\frac{1}{3}C_3+C_4+\frac{1}{3}C_9+C_{10}) F_{e,\pi K}  \nonumber\\
&& +(\frac{1}{3}C_5+C_6+\frac{1}{3}C_7+ C_{8})F^P_{e,\pi K}  +(\frac{1}{3}C_3+\frac{1}{3}C_9) M_{e,\pi K}   +(\frac{1}{3}C_5+\frac{1}{3}C_7)M^R_{e,\pi K}\nonumber\\
&&+(\frac{1}{3}C_3+C_4-\frac{1}{6}C_9-\frac{1}{2}C_{10}) F_{a,K \pi} +(\frac{1}{3}C_5+C_6+\frac{1}{3}C_7+ C_{8})F^P_{a,K \pi}\nonumber\\
&&+(\frac{1}{3}C_3-\frac{1}{6}C_9) M_{a,K \pi}+(\frac{1}{3}C_5-\frac{1}{6}C_7)M^R_{a,K \pi}
\Big],
\end{eqnarray}

\begin{eqnarray}\label{MK0p0}
-\sqrt{2}\mathcal{M} (\bar{B^0}&&\rightarrow \bar{K^0} \pi^0) =\xi_{us}\Big[ -(C_1+\frac{1}{3}C_2)F_{e,K \pi}-\frac{1}{3}C_2M_{e,K \pi} \Big] -\xi_{ts}\Big[ - \frac{3}{2} ( -C_7 -\frac{1}{3}C_8 + C_9 + \frac{1}{3}C_{10}) F_{e,K \pi} \nonumber\\
&&  + (\frac{1}{3}C_3+C_4-\frac{1}{6}C_9-\frac{1}{2}C_{10})F_{e,\pi K}  +(\frac{1}{3}C_5+C_6-\frac{1}{6}C_7-\frac{1}{2}C_{8})F^P_{e,\pi K} -(\frac{1}{2}C_{10}) M_{e,K \pi}  \nonumber\\
&& +(\frac{1}{3}C_3-\frac{1}{6}C_9) M_{e,\pi K} +(\frac{1}{3}C_5-\frac{1}{6}C_7) M^R_{e,\pi K}- (\frac{1}{2}C_{8}) M^P_{e,K \pi} \nonumber\\
&&+ (\frac{1}{3}C_3+C_4-\frac{1}{6}C_9-\frac{1}{2}C_{10})F_{a,K \pi} + (\frac{1}{3}C_5+C_6-\frac{1}{6}C_7-\frac{1}{2}C_{8})F^P_{a,K \pi} \nonumber\\
&&+ (\frac{1}{3}C_3-\frac{1}{6}C_9)M_{a,K \pi} + (\frac{1}{3}C_5-\frac{1}{6}C_7)M^R_{a,K \pi}\Big],
\end{eqnarray}

\begin{eqnarray}\label{Mp0etaq}
2\mathcal{M} (\bar{B^0}&&\rightarrow \pi^0 \eta_q)=\xi_{ud}\Big[ (C_1+\frac{1}{3}C_2)(-F_{e,\pi \eta_q}+F_{e,\eta_q \pi}+F_{a,\pi \eta_q}+F_{a,\eta_q \pi}) +\frac{1}{3}C_2(-M_{e,\pi \eta_q}+M_{e,\eta_q \pi}+M_{a,\pi \eta_q}+M_{a,\eta_q \pi})\Big]\nonumber\\
&&-\xi_{td}\Big[  (-\frac{1}{3}C_3-C_4+\frac{5}{3}C_9+C_{10})(F_{e,\eta_q \pi} +F_{a,\eta_q \pi}+F_{a,\pi \eta_q})+ \frac{3}{2}( C_7 +\frac{1}{3}C_8)(-F_{e,\eta_q \pi} +F_{a,\pi \eta_q}+F_{a,\eta_q \pi}) \nonumber\\
&&  + (-\frac{1}{3}C_5-C_6+\frac{1}{6}C_7+\frac{1}{2}C_{8})   (F^P_{e,\pi \eta_q}+F^P_{e,\eta_q \pi}+F^P_{a,\pi \eta_q}+F^P_{a,\eta_q \pi})+(-\frac{7}{3}C_3-\frac{5}{3}C_4-\frac{1}{3}C_9+\frac{1}{3}C_{10}+2 C_5\nonumber \\
&&+\frac{2}{3}C_6+\frac{1}{2}C_7+\frac{1}{6}C_{8})F_{e,\pi \eta_q} +(-\frac{1}{3}C_3+\frac{1}{6}C_9+\frac{1}{2}C_{10})(M_{e,\eta_q \pi}+M_{a,\eta_q\pi}+M_{a,\pi \eta_q})\nonumber \\
&&+(-\frac{1}{3}C_5+\frac{1}{2}C_7)(M^R_{e,\pi \eta_q}+M^R_{e,\eta_q\pi}+M^R_{a,\pi \eta_q}+M^R_{a,\eta_q\pi})+(\frac{1}{2}C_8)(M^P_{e,\eta_q\pi}+M^P_{a,\pi \eta_q}+M^P_{a,\eta_q \pi})\nonumber \\
&&+(-\frac{1}{3}C_3-\frac{2}{3}C_4+\frac{1}{2}C_9-\frac{1}{2}C_{10}) M_{e,\pi \eta_q}+(-\frac{2}{3}C_6-\frac{1}{6}C_8)M^P_{e,\pi \eta_q} \Big]
\end{eqnarray}
\begin{eqnarray}\label{Mp0etas}
\sqrt{2}\mathcal{M} (\bar{B^0}&&\rightarrow \pi^0 \eta_s)=-\xi_{td}\Big[  (-C_3-\frac{1}{3}C_4+\frac{1}{2}C_9+\frac{1}{6}C_{10}+C_5+\frac{1}{3}C_6-\frac{1}{2}C_7-\frac{1}{6}C_{8})F_{e,\pi \eta_s}\nonumber\\
&& +(-\frac{1}{3}C_4+\frac{1}{6}C_{10})M_{e,\pi \eta_s}+(-\frac{1}{3}C_6+\frac{1}{6}C_{8})M^R_{e,\pi \eta_s} \Big],
\end{eqnarray}
\begin{eqnarray}\label{Mp-etaq}
\sqrt{2}\mathcal{M} (B^-&&\rightarrow \pi^- \eta_q)=\xi_{ud}\Big[ (\frac{1}{3}C_1+C_2)(F_{e,\eta_q \pi}+F_{a,\pi \eta_q}+F_{a,\eta_q \pi}) +(C_1+\frac{1}{3}C_2)F_{e,\pi \eta_q}+(\frac{1}{3}C_2)M_{e,\pi \eta_q}\nonumber \\
&&+(\frac{1}{3}C_1)(M_{e,\eta_q \pi}+M_{a,\pi \eta_q}+M_{a,\eta_q \pi})\Big]-\xi_{td}\Big[(\frac{7}{3}C_3+\frac{5}{3}C_4+\frac{1}{3}C_9-\frac{1}{3}C_{10}-2 C_5\nonumber \\
&&-\frac{2}{3}C_6-\frac{1}{2}C_7-\frac{1}{6}C_{8})F_{e,\pi \eta_q} + (\frac{1}{3}C_5+C_6-\frac{1}{6}C_7-\frac{1}{2}C_{8})F^P_{e,\pi \eta_q} +(\frac{1}{3}C_3+C_4+\frac{1}{3}C_9+C_{10})F_{e,\eta_q \pi} \nonumber \\
&&+ (\frac{1}{3}C_5+C_6+\frac{1}{3}C_7+C_{8})F^P_{e,\eta_q\pi} +(\frac{1}{3}C_3+\frac{2}{3}C_4-\frac{1}{6}C_9+\frac{1}{6}C_{10})M_{e,\pi \eta_q} +(\frac{1}{3}C_5-\frac{1}{6}C_7)M^R_{e,\pi \eta_q}\nonumber\\
&&+(\frac{2}{3}C_6+\frac{1}{6}C_7)M^P_{e,\pi \eta_q} +(\frac{1}{3}C_3+\frac{1}{3}C_9)M_{e,\eta_q \pi}+ (\frac{1}{3}C_5+\frac{1}{3}C_7)M^R_{e,\eta_q\pi} \nonumber \\
&&+(\frac{1}{3}C_3+C_4+\frac{1}{3}C_9+C_{10})(F_{a,\pi \eta_q}+F_{a,\eta_q \pi}) + (\frac{1}{3}C_5+C_6+\frac{1}{3}C_7+C_{8})(F^P_{a,\pi \eta_q}+F^P_{a,\eta_q \pi})\nonumber \\
&&+(\frac{1}{3}C_3+\frac{1}{3}C_9)(M_{a,\pi \eta_q}+M_{a,\eta_q \pi}) + (\frac{1}{3}C_5+\frac{1}{3}C_7)(
M^R_{a,\pi \eta_q}+M^R_{a,\eta_q \pi})  \Big],
\end{eqnarray}
\begin{eqnarray}\label{Mp-etas}
\mathcal{M} (B^-&&\rightarrow \pi^- \eta_s)=-\xi_{td}\Big[  (C_3\frac{1}{3}C_4-\frac{1}{2}C_9-\frac{1}{6}C_{10} +C_5+\frac{1}{3}C_6-\frac{1}{2}C_7-\frac{1}{6}C_{8})F_{e,\pi \eta_s}\nonumber\\
&& +(\frac{1}{3}C_4-\frac{1}{6}C_{10})M_{e,\pi \eta_s}+(\frac{1}{3}C_6-\frac{1}{6}C_{8})M^R_{e,\pi \eta_s} \Big],
\end{eqnarray}
\begin{eqnarray}\label{MK0etaq}
\sqrt{2}\mathcal{M} (\bar{B^0}&&\rightarrow \bar{K^0} \eta_q)=
\xi_{us}\Big[ (C_1+\frac{1}{3}C_2)F_{e,K \eta_q} +(\frac{1}{3}C_2)M_{e,K \eta_q}\Big]-\xi_{ts}\Big[  (\frac{1}{3}C_3+C_4-\frac{1}{6}C_9-\frac{1}{2}C_{10})F_{e,\eta_q K}\nonumber\\
&&  + (\frac{1}{3}C_5+C_6-\frac{1}{6}C_7-\frac{1}{2}C_{8})F^P_{e,\eta_q K}    + (2C_3+\frac{2}{3}C_4+\frac{1}{6}C_9+\frac{1}{2}C_{10}-2C_5\nonumber \\
&&-\frac{2}{3}C_6-\frac{1}{2}C_7-\frac{1}{6}C_{8})F_{e,K \eta_q} + (\frac{1}{3}C_3-\frac{1}{6}C_9)M_{e,\eta_q K}+(\frac{1}{3}C_5-\frac{1}{6}C_7)M^R_{e,\eta_q K}   \nonumber\\
&& + (\frac{2}{3}C_4+\frac{1}{6}C_9)M_{e,K \eta_q}+(\frac{2}{3}C_6+\frac{1}{2}C_{8})M^P_{e,K \eta_q}+ (\frac{1}{3}C_3+C_4-\frac{1}{6}C_9-\frac{1}{2}C_{10})F_{a,\eta_q K}\nonumber\\
&&  + (\frac{1}{3}C_5+C_6-\frac{1}{6}C_7-\frac{1}{2}C_{8})F^P_{a,\eta_q K}     +(\frac{1}{3}C_3-\frac{1}{6}C_9)M_{a,\eta_q K}+(\frac{1}{3}C_5-\frac{1}{6}C_7)M^R_{a,\eta_q K}  \Big],
\end{eqnarray}
\begin{eqnarray}\label{MK-etaq}
\sqrt{2}\mathcal{M} (B^-&&\rightarrow K^- \eta_q)=\xi_{us}\Big[ (\frac{1}{3}C_1+C_2)(F_{e,\eta_q \pi}+F_{a,K \eta_q}) +(\frac{1}{3}C_1+C_2)F_{e,K \eta_q}+(\frac{1}{3}C_2)M_{e,K \eta_q}+(\frac{1}{3}C_1)(M_{e,\eta_q \pi}\nonumber \\
&&+M_{a,K \eta_q})\Big]-\xi_{ts}\Big[(\frac{1}{3}C_3+C_4+\frac{1}{3}C_9+C_{10})F_{e,\eta_q K}+(\frac{1}{3}C_5+C_6+\frac{1}{3}C_7+\frac{1}{3}C_{8})F^P_{e,\eta_q K}\nonumber\\
&&+(2C_3+\frac{2}{3}C_4+\frac{1}{2}C_9+\frac{1}{6}C_{10}-2C_5 -\frac{2}{3}C_6-\frac{1}{2}C_7-\frac{1}{6}C_{8})F_{e,K \eta_q} +(\frac{1}{3}C_3+\frac{1}{3}C_9)M_{e,\eta_q K}\nonumber\\
&& +(\frac{1}{3}C_5+\frac{1}{3}C_7)M^R_{e,\eta_q K}+(\frac{2}{3}C_4+\frac{1}{6}C_{10})M_{e,K \eta_q} +(\frac{2}{3}C_6+\frac{1}{6}C_8)M^P_{e,K \eta_q}+(\frac{1}{3}C_3+C_4+\frac{1}{3}C_9+C_{10})F_{a,K \eta_q} \nonumber \\
&&+ (\frac{1}{3}C_5+C_6+\frac{1}{3}C_7+C_{8})F^P_{a,K \eta_q}+(\frac{1}{3}C_3+\frac{1}{3}C_9)M_{a,K \eta_q} + (\frac{1}{3}C_5+\frac{1}{3}C_7)
M^R_{a,K \eta_q}\Big],
\end{eqnarray}
\begin{eqnarray}\label{MK0etas}
\mathcal{M} (\bar{B^0}&&\rightarrow \bar{K^0} \eta_s)=-\xi_{ts}\Big[  (\frac{4}{3}C_3+\frac{4}{3}C_4-\frac{2}{3}C_9-\frac{2}{3}C_{10} -C_5-\frac{1}{3}C_6+\frac{1}{2}C_7+\frac{1}{6}C_{8})F_{e,K \eta_s}\nonumber\\
&&+(\frac{1}{3}C_5+C_6-\frac{1}{6}C_7-\frac{1}{2}C_8)F^P_{e,K \eta_s} +(\frac{1}{3}C_3+\frac{1}{3}C_4-\frac{1}{6}C_{9}-\frac{1}{6}C_{10})M_{e,K \eta_s}\nonumber \\
&&+(\frac{1}{3}C_5-\frac{1}{6}C_{7})M^R_{e,K \eta_s}+(\frac{1}{3}C_6-\frac{1}{6}C_{8})M^P_{e,K \eta_s} +(\frac{1}{3}C_3+C_4-\frac{1}{6}C_9-\frac{1}{2}C_{10})F_{a,\eta_q K} \nonumber\\
&& +(\frac{1}{3}C_5+C_6-\frac{1}{6}C_7-\frac{1}{2}C_8)F^P_{a,\eta_s K} +(\frac{1}{3}C_3-\frac{1}{6}C_9)M_{a,\eta_s K} +(\frac{1}{3}C_5-\frac{1}{6}C_7)M^P_{a,\eta_s K}  \Big],
\end{eqnarray}
\begin{eqnarray}\label{MK-etas}
\mathcal{M} (B^-&&\rightarrow K^- \eta_s)=\xi_{us}\Big[ (\frac{1}{3}C_1+C_2)F_{a,\eta_s K}+(\frac{1}{3}C_2)M_{a,\eta_s K}\Big]-\xi_{ts}\Big[  (\frac{4}{3}C_3+\frac{4}{3}C_4-\frac{2}{3}C_9-\frac{2}{3}C_{10} \nonumber \\
&&-C_5-\frac{1}{3}C_6+\frac{1}{2}C_7+\frac{1}{6}C_{8})F_{e,K \eta_s}+(\frac{1}{3}C_5+C_6-\frac{1}{6}C_7-\frac{1}{2}C_8)F^P_{e,K \eta_s} \nonumber \\
&& +(\frac{1}{3}C_3+\frac{1}{3}C_4-\frac{1}{6}C_{9}-\frac{1}{6}C_{10})M_{e,K \eta_s}+(\frac{1}{3}C_5-\frac{1}{6}C_{7})M^R_{e,K \eta_s}+(\frac{1}{3}C_6-\frac{1}{6}C_{8})M^P_{e,K \eta_s}\nonumber \\
&& +(\frac{1}{3}C_3+C_4+\frac{1}{3}C_9+C_{10})F_{a,\eta_s K}  +(\frac{1}{3}C_5+C_6+\frac{1}{3}C_7+C_8)F^P_{a,\eta_s K} \nonumber \\
&&+(\frac{1}{3}C_3+\frac{1}{3}C_9)M_{a,\eta_s K} +(\frac{1}{3}C_5+\frac{1}{3}C_7)M^R_{a,\eta_s K}  \Big],
\end{eqnarray}

\begin{eqnarray}\label{MK-K0}
\mathcal{M} (B^-&&\rightarrow K^- K^0 )=\xi_{ud}\Big[ (\frac{1}{3}C_1+C_2)F_{a,K \bar{K}} +(\frac{1}{3}C_1)M_{a,K \bar{K}}\Big]-\xi_{td}\Big[  (\frac{1}{3}C_3+C_4-\frac{1}{6}C_9-\frac{1}{2}C_{10})F_{e,K \bar{K}}\nonumber\\
&&  + (\frac{1}{3}C_5+C_6-\frac{1}{6}C_7-\frac{1}{2}C_{8})F^P_{e,K \bar{K}}    +(\frac{1}{3}C_3-\frac{1}{6}C_9)M_{e,K \bar{K}}+(\frac{1}{3}C_5-\frac{1}{6}C_7)M^R_{e,K \bar{K}}   \nonumber\\
&&+ (\frac{1}{3}C_3+C_4+\frac{1}{3}C_9+C_{10})F_{a,K \bar{K}} + (\frac{1}{3}C_5+C_6+\frac{1}{3}C_7+C_{8})F^P_{a,K \bar{K}}   \nonumber\\
&& +(\frac{1}{3}C_3+\frac{1}{3}C_9)M_{a,K \bar{K}}+(\frac{1}{3}C_5+\frac{1}{3}C_7)M^R_{a,K \bar{K}}  \Big],
\end{eqnarray}
\begin{eqnarray}\label{MK+K-}
\mathcal{M} (\bar{B^0}&&\rightarrow K^+ K^-)
=\xi_{ud}\Big[ (C_1+\frac{1}{3}C_2)F_{a,K \bar{K}} +(\frac{1}{3}C_2)M_{a,K \bar{K}}\Big]-\xi_{td}\Big[  (2C_3+\frac{2}{3}C_4+\frac{1}{2}C_9+\frac{1}{6}C_{10} \nonumber \\
&&+2C_5+\frac{2}{3}C_6+\frac{1}{2}C_7+\frac{1}{6}C_{8})F_{e,K \bar{K}}+(\frac{2}{3}C_4+\frac{1}{6}C_{10})M_{a,K \bar{K}}+(\frac{2}{3}C_6+\frac{1}{6}C_{8})M^P_{a,K \bar{K}} \Big],\nonumber \\
&&
\end{eqnarray}
\begin{eqnarray}\label{MK0K0}
\sqrt{2}\mathcal{M} (\bar{B^0}&&\rightarrow \bar{K^0} K^0)=-\xi_{td}\Big[(\frac{1}{3}C_3+C_4-\frac{1}{6}C_9-\frac{1}{2}C_{10})F_{e,K \bar{K}}+(\frac{1}{3}C_5+C_6-\frac{1}{6}C_7-\frac{1}{2}C_{8})F^P_{e,K \bar{K}}\nonumber\\
&&+(\frac{1}{3}C_3-\frac{1}{6}C_9)M_{e,K \bar{K}} +(\frac{1}{3}C_5-\frac{1}{6}C_7)M^R_{e,K \bar{K}}+(\frac{7}{3}C_3+\frac{5}{3}C_4-\frac{7}{6}C_9-\frac{5}{6}C_{10}+2C_5\nonumber\\
&&+\frac{2}{3}C_6-C_7-\frac{1}{3}C_{8})F_{a,K \bar{K}} + (\frac{1}{3}C_5+C_6-\frac{1}{6}C_7-\frac{1}{2}C_{8})F^P_{a,K \bar{K}}\nonumber \\
&&+(\frac{1}{3}C_3+\frac{2}{3}C_4-\frac{1}{6}C_9-\frac{1}{3}C_{10})M_{a,K \bar{K}}+(\frac{1}{3}C_5-\frac{1}{6}C_{7})M^R_{a,K \bar{K}} +(\frac{2}{3}C_6-\frac{1}{3}C_{8})M^P_{a,K \bar{K}} \Big], \nonumber \\
&&
\end{eqnarray}

\end{widetext}
where $\xi_{ud} = V_{ub}V_{ud}^{*}$, $\xi_{td}=V_{tb} V_{td}^{*},\xi_{us} = V_{ub}V_{us}^{*}$, $\xi_{ts}=V_{tb} V_{ts}^{*} $, the subscript of $F_{e,\pi K}$ means the pion is $M_1$ and the kaon is the external emitted meson $M_2$ in Fig. 1(a). The decay width is expressed as
\begin{equation}
	\Gamma(B \rightarrow f) =\frac{G_F^2 m_B^3}{128 \pi} |\mathcal{M}(B\rightarrow f)|^2.
\end{equation}

\subsection{Next-to-leading-order corrections}

Several very important NLO contributions to the $B \rightarrow PP$ decays are considered in this work. They are the vertex corrections, the quark loops, and the magnetic penguins, which have been calculated in the PQCD approach in Ref. \cite{LiMiSa2005}. The NLO corrections can be included by modifying the combinations of the Wilson coefficients defined below

\begin{eqnarray}
&&	a_1(\mu) =C_2(\mu) +\frac{C_1(\mu)}{N_c}, \nonumber\\
&&	a_2(\mu) =C_1(\mu) +\frac{C_2(\mu)}{N_c}, \nonumber\\
&&	a_i(\mu) =C_i(\mu) +\frac{C_{i\pm 1}(\mu)}{N_c}, ~~ i=3-10  
\end{eqnarray}
where the plus (minus) sign is for the case when $i$ is odd (even).

\subsubsection{Vertex correction}

The contributions of vertex corrections to the Wilson coefficients are \cite{QCDf1,QCDf2,QCDf3,LiMiSa2005}
\begin{eqnarray}
&&	a_1(\mu) \rightarrow a_1(\mu)+ \frac{\alpha_s(\mu)}{4 \pi}C_F\frac{C_1(\mu)}{N_c}V_1 ,\nonumber\\
&&	a_2(\mu) \rightarrow a_2(\mu)+ \frac{\alpha_s(\mu)}{4 \pi}C_F\frac{C_2(\mu)}{N_c}V_2 ,\\
&&	a_i(\mu) \rightarrow a_i(\mu)+ \frac{\alpha_s(\mu)}{4 \pi}C_F\frac{C_{i \pm 1}(\mu)}{N_c}V_i ,\space   i=3-10, \nonumber
\end{eqnarray}
In the naive dimensional regularization (NDR) scheme the function $V_i$ is given by  \cite{QCDf1,QCDf2,QCDf3}
\begin{widetext}

\begin{equation}
	V_i=\left\{
\begin{aligned}
	&12\ln\frac{m_b}{\mu}-18+\int_{0}^{1}dx \phi_{M}(x)g(x),& \mathrm{for} \; i =1-4,9,10, \\
	-&12\ln\frac{m_b}{\mu} + 6 -\int_{0}^{1}dx \phi_{M}(x)g(1-x),& \mathrm{for} \; i =5,7,   \\
	-&6 +\int_{0}^{1}dx \phi_P^{M}(x)h(1-x),& \mathrm{for} \; i= 6,8
\end{aligned}
\right.
\end{equation}
where $\phi_{M}(x)$ and $\phi^P_{M}(x)$ are the distribution amplitudes of twist 2 and 3 for the emitted meson, respectively. The hard kernels $g(x)$ and $h(x)$ are
\begin{equation}
	g(x) = 3 \left( \frac{1-2x}{1-x} \ln x -i \pi \right)+ \left[ 2\mathrm{Li}_2(x)- \ln^2 x-\frac{2 \ln x}{1-x} -(3+2i\pi)\ln x - (x \leftrightarrow 1 -x)  \right],
\end{equation}
\begin{equation}
	h(x) =   2\mathrm{Li}_2(x)- \ln^2 x -(1+2i\pi)\ln x - (x \leftrightarrow 1 -x)  .
\end{equation}

\end{widetext}

\subsubsection{The quark-loop contributions}

For the $b\rightarrow d(s)$ transition, the effective Hamiltonian contributed by the virtual quark loops is \cite{LiMiSa2005}

\begin{eqnarray}
	\mathcal{H}_{\mathrm{eff}} = &-&\sum_{q=u,c,t}\sum_{q'}\frac{G_F}{\sqrt{2}}V_{qb}V_{qd(s)}^*\frac{\alpha_s(\mu)}{2\pi}C^{(q)}(\mu,l^2)  \nonumber \\
	&\times& (\bar{d}(\bar{s})\gamma_{\rho}(1-\gamma_5)T^a b)(\bar{q}'\gamma^{\rho}T^a q'),
\end{eqnarray}

where the function $C^{(q)}(\mu,l^2)$ is
\begin{equation}\label{cq}
		C^{(q)}(\mu,l^2) =\left[G^{(q)}(\mu,l^2) -\frac{2}{3} \right]C_2(\mu)
\end{equation}
for $q =u,c$, while for $q=t$, the function is
\begin{eqnarray}\label{ct}
	C^{(t)}(\mu,l^2) =&&\left[  G^{(d)}(\mu,l^2) -\frac{2}{3}\right]C_3(\mu) \nonumber \\
		&+ & \sum_{q''=u,d,s,c}G^{(q'')}(\mu,l^2)\left[C_4(\mu) +C_6(\mu) \right]. \nonumber \\
\end{eqnarray}
The function $G$ in Eqs. (\ref{cq}) and (\ref{ct}) is
\begin{equation}
	 G^{(q)}(\mu,l^2) = -4 \int_{0}^{1}dx x(1-x) \ln \frac{m_q^2-x(1-x)l^2-i\varepsilon}{\mu^2},
\end{equation}
where $m_q$ is the quark mass for $q=u,d,s,c$.

The topology of the quark-loop contribution to the effective Hamiltonian is just the same as that of the penguin diagram, so its contribution can be absorbed into the Wilson coefficients $a_4,a_6 $
\begin{equation} \label{l-square}
	a_{4,6}(\mu) \rightarrow a_{4,6}(\mu)+\frac{\alpha_s(\mu)}{9\pi}\sum_{q=u,c,t}\frac{V_{qb}V_{	qd}^*}{V_{tb}V_{td}^*}C^{(q)}(\mu,\left<l^2\right>).
\end{equation}
$\left<l^2\right>$ in Eq. (\ref{l-square}) is the mean value of the momentum squared of the virtual gluon connecting the virtual quark loop and the final quark-antiquark pair. $\left<l^2\right>=m_b^2/4$ can be taken in the numerical analysis as a reasonable value in $B$ decays.

\subsubsection{Magnetic penguins}

The effective Hamiltonian of the magnetic penguin for the weak $ b \rightarrow d(s)\textsl{g} $ transition is
\begin{equation}\label{H-mpg}
	\mathcal{H}_{\mathrm{eff}}= - \frac{G_F}{\sqrt{2}}V_{tb}V_{td(s)}^*C_{8\textsl{g}}O_{8\textsl{g}},
\end{equation}
where the magnetic-penguin operator is
\begin{equation}
	O_{8\textsl{g}}=\frac{g}{8 \pi^2}m_b \bar{d}_i(\bar{s}_i) \sigma_{\mu \nu}(1+\gamma_5)T_{ij}^a G^{a\mu\nu} b_j.
\end{equation}
The contribution of the Hamiltonian in Eq. (\ref{H-mpg}) can be absorbed into the relevant Wilson coefficients \cite{LiMiSa2005}
\begin{equation}
	a_{4,6}(\mu) \rightarrow a_{4,6}(\mu)-\frac{\alpha_s(\mu)}{9\pi} \frac{2m_B}{\sqrt{\left<l^2\right>}}C_{8\textsl{g}}^{\mathrm{eff}}(\mu),
\end{equation}
where the effective coefficient $C_{8\textsl{g}}^{\mathrm{eff}}=C_{8\textsl{g}}+C_5$ \cite{Hamiltanion1996}.

\subsubsection{Spectator hard-scattering mechanism with $g*g* \rightarrow \eta(\eta')$}

\begin{figure}[bth]
	\epsfig{file=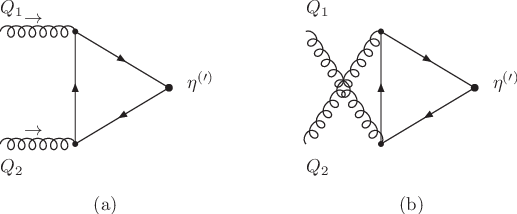,width=7.2cm,height=3.0cm}
	\caption{The Feynman diagrams for the effective interaction of $g*g* \rightarrow \eta(\eta')$ transition, where diagrams (a) and (b) represent two distinct structures.} \label{fig2}
\end{figure}

There is the contribution of the spectator hard-scattering mechanism (SHSM) for processes of $\eta(\eta')$ production through the transition of $g*g* \rightarrow \eta(\eta')$ \cite{DKY1998,AKS1998,DuY1998,MutaY2000,YY2001}. It may significantly enhance the branching ratios of $B$ decays involving $\eta(\eta')$ in the final states. In this work we incorporate this mechanism in the calculation of the amplitude for the processes with $\eta$ or $\eta'$ in the final state. The difference from the previous works is that the transverse momenta of the quarks and gluons are included in the calculations, both in the effective transition of $g*g* \rightarrow \eta(\eta')$ and the spectator hard scattering. The diagrams for the $g*g* \rightarrow \eta(\eta')$ transition are depicted in Fig. \ref{fig2}.

\begin{figure}[bth]
	\epsfig{file=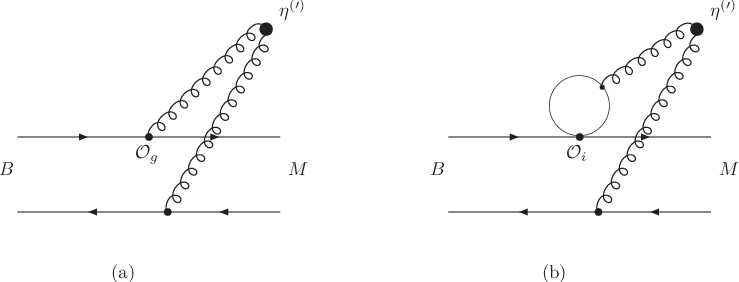,width=8.2cm,height=3.6cm}
	 \caption{The Feynman diagrams for the spectator hard-scattering contribution to $B\rightarrow M \eta^{(\prime)}$ decays, with diagrams (a) and (b) depicting contributions from the magnetic-penguin operator and the quark-loop process, respectively. The solid dots stand for the effective vertex of $g*g* \rightarrow \eta(\eta')$ transition as shown in Fig. \ref{fig2}.} \label{fig3}
\end{figure}
The SHSM includes two types of contributions. One is induced by the magnetic-penguin operator, while the other arises from the quark-loops process. The diagrams are shown in Fig. \ref{fig3}.

The amplitude for the contribution of the magnetic-penguin operator [Fig. \ref{fig3} (a)] is

\begin{eqnarray}\label{MO8g}
	&\mathcal{M}_{O_{8g}} &=-V^*_{tq}V_{tb}C_{8\textsl{g}}^{\mathrm{eff}}(\mu)(2f^u_{\eta^{(')}}+f^s_{\eta^{(')}})\mathcal{F}_{O_{8g}},
\end{eqnarray}
where
\begin{eqnarray}\label{FO8g}
&\mathcal{F}_{O_{8g}} &=-i\frac{m_B^3}{ N_c^3}f_B f_{M_1} \int dk_{\perp}k_{\perp}\int_{x^d}^{x^u}dx\int_0^1 dx_1 dx_2 \nonumber\\
&&\int_0^\infty bdbb_2db_2\int_0^{2 \pi} d\theta\int_0^1 du (\frac{1}{2}m_B+\frac{|\vec{k}_{\perp}|^2}{2x^2m_B})\nonumber\\
&&\times K(\vec{k})(E_Q+m_Q)J_0(k_{\perp}b) \Bigg\{\alpha_s^2(\mu_{1}) x_1\nonumber\\
&&\times \phi_{\eta_{q(s)}}(\bar{x}_2,b_2)\Bigg(m_B[E_q-k^3]\phi_{M_1}(\bar{x}_1,b)\nonumber\\
&& +\mu_{M_1}[E_q(x_1-1)-k^3(x_1+1)]\phi^P_{M_1}(\bar{x}_1,b)\nonumber\\
&&+\frac{1}{6}\mu_{M_1}[E_q(1 + x_1)+k^3(1-x_1)]\phi'^\sigma_{M_1}(\bar{x}_1,b)\Bigg)\nonumber\\
&&\times h_1(x,x_1,x_2,b,b_2,\theta,u)S_t(x_1) \nonumber\\
&& \times \exp[-S_{B}(\mu_{1})-S_{M_1}(\mu_{1})-S_{M_2}(\mu_{1})]\Bigg\}.
\end{eqnarray}

The contribution of the second type [Fig. \ref{fig3} (b)] is
\begin{equation}\label{MQL}
	\mathcal{M}_{ql} =\sum_{f=u,c,t}V^*_{fq}V_{fb} C^{(f)}(\mu,l^2) (2f^u_{\eta^{(')}}+f^s_{\eta^{(')}})\mathcal{F}_{ql}
\end{equation}
where
\begin{eqnarray}\label{FQL}
&\mathcal{F}_{ql} &=-i\frac{2m_B}{N_c^3}f_B f_{M_1}  \int dk_{\perp}k_{\perp}\int_{x^d}^{x^u}dx\int_0^1 dx_1 dx_2 \nonumber\\
&&\int_0^\infty bdbb_2db_2(\frac{1}{2}m_B+\frac{|\vec{k}_{\perp}|^2}{2x^2m_B})K(\vec{k})(E_Q+m_Q)\nonumber\\
&&\times J_0(k_{\perp}b) \alpha_s^2(\mu_{1}) x_1 \phi_{\eta_{q(s)}}(\bar{x}_2,b_2)\Bigg(\frac{1}{2}m_B(E_q-k^3)\nonumber\\
&&\times \phi_{M_1}(\bar{x}_1,b)-\mu_{M_1}k^3\phi^P_{M_1}(\bar{x}_1,b)+\frac{1}{6}\mu_{M_1}E_q\nonumber
\end{eqnarray}	
\begin{eqnarray}
&&\times \phi'^\sigma_{M_1}(\bar{x}_1,b)\Bigg)h_d^2(x,x_1,x_2,b,b_2)S_t(x_1) \nonumber\\
&& \times \exp[-S_{B}(\mu_{1})-S_{M_1}(\mu_{1})-S_{M_1}(\mu_{1})].
\end{eqnarray}

In Eqs. (\ref{FO8g}) and (\ref{FQL}), the function $h_1(x,x_1,x_2,b,b_2,\theta,u)$ and the scale $\mu_{1}$ are
\begin{widetext}
	
	\begin{equation}
		h_1(x,x_1,x_2,b,b_2,\theta,u) =\left\{
		\begin{aligned}
			\frac{\sqrt{b^2+b_2^2+2b b_2 \cos(\theta)}}{2\sqrt{(x-u)x_1}m_B}&K_{-1}(\sqrt{b^2+b_2^2+2b b_2 \cos(\theta)}\sqrt{(x-u)x_1}m_B)&\\
			\cdot &K_0(-i\sqrt{x_1x_2}m_B b_2),& \mathrm{for} \;u < x, \\
			&&\\
			i\frac{\sqrt{b^2+b_2^2+2b b_2 \cos(\theta)}}{2\sqrt{(u-x)x_1}m_B}&K_{-1}(-i\sqrt{b^2+b_2^2+2b b_2 \cos(\theta)}\sqrt{(u-x)x_1}m_B)& \\
			\cdot &K_0(-i\sqrt{x_1x_2}m_B b_2),& \mathrm{for} \;u > x,
		\end{aligned}
		\right.
	\end{equation}

	\begin{equation}
		\mu_1 = \max(\sqrt{x_1}m_B,\sqrt{x_1x_2}m_B,1/b,1/b_2)
	\end{equation}
\end{widetext}

\vspace{0.5em}

\section{The Contribution of Soft Form Factors}

As studied in Refs. \cite{Lu-Yang2021,lu-yang2023,wang-yang2023}, large soft contributions still exist in diagrams (a), (b), (g), and (h) in Fig. \ref{fig1} as the $B$ meson wave function solved from the relativistic potential model being used. To keep the perturbative calculation reliable, a cutoff scale $\mu_c$ needs to be introduced. For contributions with scale $\mu >\mu_c$, they can be calculated by the perturbative QCD method, while the contributions with scale $\mu <\mu_c$
are replaced by two kinds of soft form factors, the soft $BM$ transition form factor and the $M_1M_2$ production form factor, where $M$, $M_1$, and $M_2$ denote mesons in the final state of $B$ decays. In general, the critical cutoff scale can be taken as $\mu_c=1\;\mathrm{GeV}$. As $\mu_c$ slightly varied around 1 GeV, the physical results of branching ratios and $CP$ violations are not changed much \cite{lu-yang2023,wang-yang2023}. For Figs. \ref{fig1}(c) and \ref{fig1}(d), contributions are still dominated by the perturbative contribution with $\alpha_s/\pi <0.2$. In general, the contributions of nonfactorizable annihilation diagrams (e) and (f) in Fig. \ref{fig1} are very small. Therefore, for decay modes where the contributions of nonfactorizable annihilation diagrams are small, no soft contributions need to be introduced for them.

The total $BM$ transition form factor can be separated into two parts
\begin{equation} \label{softBM}
F_0^{BM}=h_0^{BM}+\xi^{BM},
\end{equation}
where $h_0^{BM}$ is the hard $BM$ transition form factor, which is contributed by hard interaction, and $\xi^{BM}$ the soft part of the transition form factor. Including the contributions of the soft transition form factor, the amplitude relevant to diagrams (a) and (b) in Fig. \ref{fig1} is changed as
\begin{eqnarray}
	\mathcal{M} \rightarrow \mathcal{M}&-&2if_{\pi}C(\mu_{c})V_{\mathrm{CKM}} \cdot\xi^{BM}\nonumber\\
	&-&4i\frac{\mu_M}{m_B}f_{M}C'(\mu_{c})V_{\mathrm{CKM}} \cdot\xi^{BM},
\end{eqnarray}
where $C(\mu_{c})$ and $C'(\mu_{c})$ are the relevant Wilson coefficients for the operators of $(V-A)(V-A)$ and $(S+P)(S-P)$ at the critical cutoff scale $\mu_{c}$, respectively.

The soft contributions stemming from the factorizable annihilation diagrams (g) and (h) in Fig.\ref{fig1} can be absorbed into the soft production form factor of $M_1M_2$. The soft $M_1M_2$ production form factor can be defined by the matrix element of the scalar current as
\begin{equation}\label{s-formfactor}
	\langle M_1M_2|S|0\rangle =-\dfrac{1}{2}\sqrt{\mu_{M_1}\mu_{M_2}} F_{+}^{M_1M_2}(q^2),
\end{equation}
where $\mu_{M}=m_M^2/(m_{q_1}+m_{q_2})$ is the chiral parameter for the charged meson,  $M=M_1$ or $M_2$, and $q_{1,2}$ the quark-antiquark in the meson $M_1$ or $M_2$. The form factor $F_{+}^{\pi\pi}$ can also be separated into two parts, the hard and soft parts
\begin{equation} \label{s-form-hard-and-soft}
F_{+}^{M_1M_2}=h^{M_1M_2}+\xi^{M_1M_2},
\end{equation}
where $h^{M_1M_2}$ is the hard production form factor for $M_1M_2$, and $\xi^{M_1M_2}$ the soft part of the form factor. The soft form factor contributes to the amplitude as
\begin{equation}
	\mathcal{M} \rightarrow \mathcal{M}+\frac{2\sqrt{\mu_{M_1}\mu_{M_2}}}{m_B^2} \langle0|S-P|B\rangle C(\mu_{c})V_{\mathrm{CKM}} \xi^{M_1M_2},
\end{equation}
where $\langle0|S-P|B\rangle=-i\chi_B$, and $\chi_B$ can be found in Eq. (\ref{chiB}).

\section{color-octet contribution}
The contributions of color-octet quark-antiquark pair in the final state are usually dropped in the theoretical calculation if the quark-antiquark pair finally forms one meson in the decay process, because mesons should be in color singlet. However, the contributions of the diagrams in Fig. \ref{fig1} with the quark-antiquark pair in the final state being in color octet may not be zero. In principle, the quark-antiquark pairs in the final state of Fig. \ref{fig1} can be produced in color-octet states after short-distance interaction. As the color-octet quark pairs move away from each other to the hadron scale, they can finally be changed into color-singlet states by exchanging soft gluons. Therefore the color-octet quark-antiquark pairs can contribute to the decay process of the $B$ meson. This mechanism has been introduced by us to solve the $\pi\pi$ and $K\pi$ puzzles in $B$ decays recently \cite{lu-yang2023,wang-yang2023}. In this work we extend this mechanism to more decay modes with two pseudoscalar mesons in the final state.

\begin{figure}[htb]
  \includegraphics[width=0.40\textwidth]{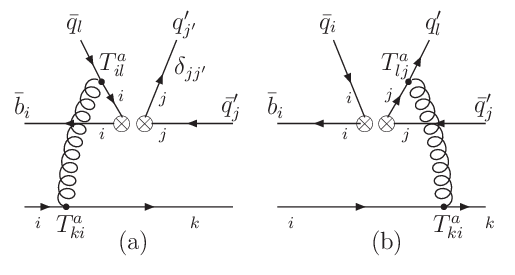}
  \caption{\label{fig-color8} Two nonfactorizable diagrams with an operator insertion of $(\bar{b}_iq_i)(\bar{q}^\prime_j q^\prime_j)$, where the explicit type of the current is omitted. The quark-antiquark pairs in the final state are in nonsinglet color states. The symbols $i$, $j$, $j^\prime$, $k$, and $l$ are color indices. (a) is for the diagram where the gluon connecting the antiquark line in the upper emitted meson and the light quark line between $B$ meson and the other light meson, and (b) for the gluon connecting the light quark line in the upper emitted meson and the light quark line between $B$ and the lower emitted meson.}
\end{figure}

The details of the calculation of the color-octet contribution have been given in Ref. \cite{wang-yang2023}. Here we briefly present the main steps in this paper. To consider the color-octet contributions, we need to consider the case that the quark-antiquark pairs in Fig. \ref{fig1} that finally form the mesons in the final state are in nonsinglet state. Then one can separate the contributions of the color-octet state from the color-singlet state by analyzing the color factors appearing in each diagram in Fig. \ref{fig1}. Figure \ref{fig-color8} is an example for the treatment of the color factors, where the insertion of the operator $(\bar{b}_iq_i)(\bar{q}^\prime_j q^\prime_j)$ is considered. Operators with other color structures can be considered similarly. The color factor for Fig.~\ref{fig-color8} (a) becomes
\begin{equation}\label{eq:cfa}
  \begin{split}
    \sum_{ijkl} T_{ki}^a T_{il}^a &= \sum_{jkl} C_F \delta_{lk}
      = \sum_{jklj^\prime} C_F \delta_{lk} \delta_{jj^\prime} \\
    & = \sum_{jklj^\prime} C_F \left(\dfrac{1}{N_c} \delta_{lj^\prime} \delta_{jk} + 2 T_{lj^\prime}^a T_{jk}^a\right),
  \end{split}
\end{equation}
and the color factor for Fig.~\ref{fig-color8} (b) is
\begin{equation}\label{eq:cfb}
  \begin{split}
    &\sum_{ijkl} T_{lj}^a T_{ki}^a
    = \sum_{ijkl} \left[-\dfrac{1}{2N_c} \delta_{lj} \delta_{ki} + \dfrac{1}{2} \delta_{li} \delta_{kj} \right] \\
  &\quad = \sum_{ijkl} \left[-\dfrac{1}{2N_c} \left(\dfrac{1}{N_c} \delta_{li} \delta_{kj}
      + 2 T_{li}^b T_{kj}^b\right) + \dfrac{1}{2} \delta_{li} \delta_{kj} \right]\\
    &\quad = \sum_{ijkl} \left(\dfrac{C_F}{N_c} \delta_{li} \delta_{kj}
      - \dfrac{1}{N_c} T_{li}^b T_{kj}^b \right),
  \end{split}
\end{equation}
where the first terms with two delta functions in Eqs. (\ref{eq:cfa}) and (\ref{eq:cfb}) correspond to the color-singlet contributions, which give $M_e$ and $M_e^P$ for the nonfactorizable diagrams in Fig. \ref{fig1}, and the second terms with SU(3)$_\mathrm{c}$ generators give the color-octet contributions. The parameters that describe the nonperturbative effects where the color-octet quark-antiquark pairs are changed to color-singlet states by exchanging soft gluons need to be introduced. In numerical analysis we find that two parameters $Y^8_F$ and $Y^8_M $ are needed to explain the experimental data. $Y^8_F$ and $Y^8_M $ correspond to factorizable and nonfactorizable diagrams in Fig. \ref{fig1}, respectively. For diagrams (a) and (b) in Fig. \ref{fig-color8}, the result is
\begin{equation}\label{nfMe8}
 Y^8_M \mathcal{M}_e^{(P,R)8},
\end{equation}
where
\begin{equation}\label{eq:fe8}
  \mathcal{M}_e^{(P,R)8} \equiv 2N_c^2 \mathcal{M}_e^{(P,R)c} - \dfrac{N_c}{C_F} \mathcal{M}_e^{(P,R)d}.
\end{equation}
The symbols with and without the superscript $P$ denote the results for $(S+P)(S-P)$ and $(V-A)(V-A)$ operators, respectively.

The color-octet contributions for the other diagrams in Fig. \ref{fig1} with all kinds of operator insertions are
\begin{equation}
Y^8_F F_e^{(P)8},\;
Y^8_M\mathcal{M}_e^{(P,R)\prime 8},\;Y^8_M\mathcal{M}_a^{(P,R)8},\; Y^8_F F_a^{(P)8},
\end{equation}
where
\begin{equation}\label{fa8}
  \begin{split}
    & F_e^{(P)8} \equiv 2N_c^2 F_e^{(P)a} - \dfrac{N_c}{C_F} F_e^{(P)b}, \\
    &\mathcal{M}_e^{(P,R)\prime 8} \equiv \dfrac{N_c^2}{C_F} \mathcal{M}_e^{(P,R)}, \
      \mathcal{M}_a^{(P,R)8} \equiv -\dfrac{N_c}{C_F} \mathcal{M}_a^{(P,R)}\\
      &      F_a^{(P)8} \equiv -\dfrac{N_c^2}{C_F} F_a^{(P)}.
  \end{split}
\end{equation}
The quantities $F_e^{(P)a}$, $F_e^{(P)b}$, $\mathcal{M}_e^{(P,R)c}$, $\mathcal{M}_e^{(P,R)d}$,  $\mathcal{M}_a^{(P,R)}$, and $F_a^{(P)}$ are the convolution functions corresponding to diagrams (a)$\--$(h) in Fig.~\ref{fig1} by using the PQCD approach, where the distribution functions of the quark-antiquarks in color-octet states are assumed to be the same as the color-singlet states.

\section{Analysis of the Soft Parameters Under SU(3) Flavor Symmetry and Its Breaking}

\subsection{The color-octet parameters $Y^8_{F,M}$ }

The color-octet parameters $Y^8_{F}$ and $Y^8_{M}$ are long-distance parameters, which may depend on the different mesons in the final state, such as $\pi\pi$, $\pi K$ and $\pi\eta$ final states, etc. These parameters for different final states can be related by SU(3) flavor symmetry and the symmetry-breaking effect.

In the limit of SU(3) symmetry, light pseudoscalar mesons can be composed into a nonet \cite{Hexg2014,Hexg2020}
\begin{eqnarray}
M&=&\left(\begin{array}{ccc}
\frac{\pi^0}{\sqrt{2}}+\frac{\eta_8}{\sqrt{6}} +\frac{\eta_1}{\sqrt{3}}& \pi^{+} & K^{+} \\
\pi^{-} & -\frac{\pi^0}{\sqrt{2}}+\frac{\eta_8}{\sqrt{6}} +\frac{\eta_1}{\sqrt{3}} & K^0 \\
K^{-} & \bar{{K}^0}  & -\frac{2\eta_8}{\sqrt{6}} +\frac{\eta_1}{\sqrt{3}}
\end{array}\right) \nonumber\\
&=&\left(\begin{array}{ccc}
\frac{\pi^0+\eta_q}{\sqrt{2}} & \pi^{+} & K^{+} \\
\pi^{-} & \frac{-\pi^0+\eta_q}{\sqrt{2}} & K^0 \\
K^{-} & \bar{{K}^0}  & \eta_s
\end{array}\right) \label{M-nonet}
\end{eqnarray}
where the mixing of the flavor octet and singlet is considered and included in the nonet.

The color-octet parameters $Y^8_{F}$ and $Y^8_{M}$ describe the effect of two color-octet quark-antiquark pairs
$ M_8^1$ and $M_8^2$ scattering into color-singlet states $M_1^1$ and $M_1^2 $ by long-distance QCD interactions, where $M_{8,1}^{1,2}$ denote the first and second mesons in color-octet and singlet states which can be one of the matrix elements in Eq. (\ref{M-nonet}).  For the scattering of $ M_8^1 M_8^2 \rightarrow M_1^1 M_1^2 $, the effective Hamiltonian under SU(3) flavor symmetry can be written as
\begin{equation}
H_0= c_0(M_8^1)^i_j (M_8^2)^k_l \cdot ( M_1^{1\dagger})^j_i( M_1^{2\dagger})^l_k
\end{equation}
where $c_0$ is the effective coupling describing the scattering.

The SU(3) symmetry-breaking effect is caused by the large mass of $s$ quark which is apparently different from that of $u$ and $d$ quarks. A diagonal matrix $W$ is used to describe the SU(3) symmetry-breaking effect \cite{Hexg2014,Hexg2020,Wangrm2023},
\begin{equation}
W=\left(W_j^i\right)=\left(\begin{array}{ccc}
0 & 0 & 0 \\
0 & 0 & 0 \\
0 & 0 & 1
\end{array}\right).
\end{equation}

The leading order SU(3) symmetry-breaking terms are
\begin{eqnarray} \label{SU3-braaking_L}
&&H^1_1 = c^1_1 \left(W^i_j(M_8^1)^j_m  ( M_1^{1\dagger})^m_i\right) \left( (M_8^2)^k_l( M_1^{2\dagger})^l_k \right)  \nonumber \\
&&H^2_1 = c^2_1 \left((M_8^1)^i_j W^j_m ( M_1^{1\dagger})^m_i\right) \left( (M_8^2)^k_l( M_1^{2\dagger})^l_k \right)
\end{eqnarray}
where all the nonequivalent possibilities of putting the matrix $W$ in the SU(3) symmetric effective Hamiltonian should be considered. Note $\left((M_8^1)^i_j ( M_1^{1\dagger})^j_m W^m_i\right)=\left(W^i_j(M_8^1)^j_m  ( M_1^{1\dagger})^m_i\right)$, and putting $W$ in the term $\left( (M_8^2)^k_l( M_1^{2\dagger})^l_k \right)$ is finally the same as that putting $W$ in $\left((M_8^1)^i_j  ( M_1^{1\dagger})^j_i\right)$. So there are only two different ways for putting $W$ in the effective Hamiltonian at leading order of SU(3) symmetry breaking given in Eq. (\ref{SU3-braaking_L}). And $c^1_1$ and $c^2_1$ are the parameters for the leading-order SU(3) breaking effect.

Substitute the matrix $M$ in Eq. (\ref{M-nonet}) into $M_8^1$ and $M_1^1$, one can obtain
\begin{eqnarray}
&&\left(W^i_j(M_8^1)^j_m  ( M_1^{1\dagger})^m_i\right)\nonumber\\
&&=(\bar{K}^0_1)^\dagger (\bar{K}^0_8)+ (K^{-}_1)^\dagger K^{-}_8+\eta_{s1}^\dagger \eta_{s8}, \label{c11}
\end{eqnarray}

\begin{eqnarray}
&&\left((M_8^1)^i_j W^j_m ( M_1^{1\dagger})^m_i\right)\nonumber\\
&&=(K^0_1)^\dagger K^0_8+ (K^{+}_1)^\dagger K^{+}_8+\eta_{s1}^\dagger  \eta_{s8}. \label{c12}
\end{eqnarray}
The above results show that Eqs. (\ref{c11}) and (\ref{c12}) are $CP$ conjugate terms. $CP$ symmetry in strong interaction requires $c^1_1 = c^2_1$ .

The SU(3) symmetry-breaking terms at next-to-leading order should be
\begin{eqnarray}
&&H^1_2=	c^1_2 \left(W^i_j(M_8^1)^j_m ( M_1^{1\dagger})^m_i\right) \left( W^k_l(M_8^2)^l_n( M_1^{2\dagger})^n_k \right),  \nonumber \\
&&H^2_2=	c^2_2 \left(W^i_j(M_8^1)^j_m ( M_1^{1\dagger})^m_i\right) \left( (M_8^2)^k_lW^l_n( M_1^{2\dagger})^n_k \right),  \nonumber \\
&&H^3_2=	c^3_2\left((M_8^1)^i_jW^j_m ( M_1^{1\dagger})^m_i\right) \left( (M_8^2)^k_lW^l_n( M_1^{2\dagger})^n_k \right), \nonumber \\
&&H^4_2=	c^4_2 \left(W^i_j(M_8^1)^j_m W^m_n ( M_1^{1\dagger})^n_i \right) \left( (M_8^2)^k_l( M_1^{2\dagger})^l_k \right).\nonumber \\
&&	
\end{eqnarray}
Similarly, $CP$ symmetry requires $c^1_2 = c^3_2$.

Based on the analysis of SU(3) flavor symmetry, we can express the color-octet parameters $Y^8_F$ and $Y^8_M$ in terms of the SU(3) symmetry and symmetry-breaking parameters $c_0$, $c^1_1$, $c^1_2$, $c^2_2$, and $c^4_2$. An extra superscript $a$ or $b$ should be added to these parameter $c$'s to sign the difference between $Y^8_F$ and $Y^8_M$, where $a$ is for $Y^8_F$, and $b$ for $Y^8_M$. The results for different decay modes are given in Table \ref{Y8cab}.

	\begin{table*}[htb]
		\renewcommand\arraystretch{1.5}
		\caption{\label{Y8cab} The coefficients of LO and NLO SU(3) symmetry-breaking parameters for different final states, where $R_j^{ia(b)}=c_j^{ia(b)}/c_0^{a(b)}$, $i\; j=1, 2, 4$.}
		\begin{threeparttable}
			\setlength{\tabcolsep}{12mm}{
				\begin{tabular}{c|c|c}
					\hline
					\hline
		  &$Y^8_F$  &   $Y^8_M$        \\ \hline
		$\pi \pi(\eta_q)$ & $c_0^a$ &  $c_0^b$   \\
		$K \pi(\eta_q)$  &$c_0^a(1+\frac{1}{2}R_1^{1a})$  &  $c_0^b(1+\frac{1}{2}R_1^{1b})$ 	   \\
		$\pi \eta_s$ & $c_0^a(1+R_1^{1a}+\frac{1}{2}R_2^{4a})$ & $c_0^a(1+R_1^{1a}+\frac{1}{2}R_2^{4a})$  	   \\
		$K \eta_s$     & $c_0^a(1+\frac{3}{2}R_1^{1a}+R_2^{1a}+\frac{1}{2}R_2^{2a}+\frac{1}{2}R_2^{4a})$  &  $c_0^b(1+\frac{3}{2}R_1^{1b}+R_2^{1b}+\frac{1}{2}R_2^{2b}+\frac{1}{2}R_2^{4b})$       \\
		$K \bar{K}$     &  $c_0^a(1+R_1^{1a}+\frac{1}{2}R_2^{2a})$  & $c_0^b(1+R_1^{1b}+\frac{1}{2}R_2^{2b})$      \\
					\hline
					\hline
			\end{tabular}  }
		\end{threeparttable}
		
	\end{table*}

\subsection{The Production Form Factors Defined By The Matrix Element $\langle M_1M_2|S|0\rangle$}

The production form factor is defined by the matrix element induced by the scalar current in Eq. (\ref{s-formfactor}). The effective Hamiltonian for the matrix element of the scalar current in the limit of SU(3) symmetry can be written as
\begin{equation} \label{MM0}
H_0=c_0^c M^i_jM^j_k S^k_i
\end{equation}
where $M^i_j$ and $M^j_k$ are the meson states given in the SU(3) symmetric pseudoscalar nonet in Eq. (\ref{M-nonet}), $i,j,k=1,2,3$, and $S^k_i=\bar{q}_{i}q'_{k}$, the scalar current composed of quark fields. The quark fields are denoted as $q_1(q'_1)=u$, $q_2(q'_2)=d$, and $q_3(q'_3)=s$.

The leading-order symmetry-breaking terms are
\begin{eqnarray}
H_1^1&=&c_1^{1c} W^i_l M^l_jM^j_k S^k_i, \nonumber\\
H_1^2&=&c_1^{2c} M^i_j W^j_l M^l_k S^k_i, \\ \label{MM1}
H_1^3&=&c_1^{3c} M^i_j M^j_k W^k_l S^l_i. \nonumber
\end{eqnarray}
$CP$ conservation of strong interaction leads to $c_1^{1c}=c_1^{3c}$.

The second order symmetry-breaking terms are
\begin{eqnarray}
H_2^1&=&c_2^{1c} W^i_j M^j_lW^l_m M^m_k S^k_i, \nonumber\\
H_2^2&=&c_2^{2c} W^i_j M^j_l M^l_m W^m_k S^k_i, \\ \label{MM2}
H_2^3&=&c_2^{3c} M^i_j W^j_l M^l_m W^m_k S^k_i. \nonumber
\end{eqnarray}
$CP$ symmetry requires $c_2^{1c}=c_2^{3c}$.

Substituting the SU(3) symmetry nonet and symmetry-breaking matrices $M$ and $W$ into Eqs. (\ref{MM0}), (\ref{MM1}), and (\ref{MM2}), one can get the expressions of the production form factors in terms of the symmetry and symmetry-breaking parameters. The results are collected in Table \ref{S-Form}. The soft part of the production form factor can be obtained by using Eq. (\ref{s-form-hard-and-soft}), where the hard part can be calculated perturbatively.

	\begin{table*}[htb]
		\renewcommand\arraystretch{1.5}
		\caption{\label{S-Form} The coefficients of LO and NLO SU(3) symmetry-breaking parameters for production form factors with different final states, where $R_j^{ic}=c_j^{ic}/c_0^{c}$, $i\; j=1, 2, 3, 4$.}
		\begin{threeparttable}
			\setlength{\tabcolsep}{12mm}{
				\begin{tabular}{c|c}
					\hline
					\hline
		  &$\sqrt{\mu_{M_1}\mu_{M_2}} F_+^{M_1 M_2}$        \\ \hline
		$\pi \pi(\eta_q)$ & $c_0^c$    \\
		$K \pi(\eta_q)$  &$c_0^c(1+R_1^{1c})$ 	   \\
		$\pi \eta_s$ & 0  	   \\
		$K \eta_s$     & $c_0^c(1+R_1^{1c}+R_1^{2c}+R_2^{1c})$       \\
		$K \bar{K}$     &  $c_0^c(1+R_1^{2c})$       \\
					\hline
					\hline
			\end{tabular}  }
		\end{threeparttable}
		
	\end{table*}

\section{Numerical analysis and discussion}

In numerical calculations, the input parameters involve the nonperturbative parameters, including the soft $B M$ transition form factor $\xi^{BM}$, soft $M_1M_2$ production form factor $\xi^{M_1M_2}$, and the color-octet parameters $Y^8_F , Y^8_M$, except for the parameters in $B$ and light meson wave functions. The color-octet parameters are expressed as the SU(3) flavor symmetry and symmetry-breaking parameters. Among these, the determination of the soft $B M$ transition form factor requires a combined analysis of perturbative calculations and experimental data from $B$ meson semileptonic decays. The other parameters will be obtained by fitting the branching ratios and direct $CP$ violation of $ B \rightarrow M_1 M_2$ decays.

The hard part of the $BM$ transition form factors can be obtained by calculating diagrams (a) and (b) in Fig. 1 where the contribution for the emitted meson $M_2$ is excluded. The results we obtain are
\begin{eqnarray}
	&&h_+^{B\pi} = 0.23 \pm0.01,  \nonumber \\
	&&h_+^{B K} = 0.29 \pm0.02,    \\
	&&h_+^{B\eta_q} = 0.17 \pm0.01,  \nonumber
\end{eqnarray}
where the cutoff scale $\mu_c=1\;\mbox{GeV}$ is used. The scale for the hard contribution is $\mu>\mu_c$.

Based on the experimental data of $B$ meson semileptonic decays and nonperturbative methods such as light-cone sum rules and lattice QCD (LQCD) \cite{Bailey2016,PBD2023,ball2005new}, we can extract the total $BM$ transition form factors
\begin{eqnarray}
	&&F_+^{B\pi} = 0.27 \pm0.02,  \nonumber \\
	&&F_+^{B K} = 0.33 \pm0.04,    \\
	&&F_+^{B\eta_q} = 0.23 \pm 0.03. \nonumber
\end{eqnarray}
For the form factors of $F_+^{B\pi}$ and $F_+^{B\eta_q}$, the values in the above equations can be used to calculate the branching ratios of the relevant semileptonic decays, and with which the results consistent with experimental data in PDG \cite{PDG2022} can be obtained. The value of $F_+^{B K}$ is the averaged results of LQCD calculations \cite{Bailey2016,PBD2023}.

According to Eq. (\ref{softBM}), the soft part of the $BM$ transition form factors are
\begin{eqnarray}
	&&\xi^{B\pi} = 0.04 \pm0.01,  \nonumber \\
	&&\xi^{B K} = 0.04 \pm0.02,    \\
	&&\xi^{B\eta_q} = 0.06 \pm0.02 . \nonumber
\end{eqnarray}

\newcolumntype{H}{>{\setbox0=\hbox\bgroup}c<{\egroup}@{}}		
	\begin{table*}[htb]
	\renewcommand\arraystretch{1.5}
	\caption{\label{table-BRCP-I} Branching ratio ( $\times 10^{-6}$) and direct $CP$ violation with NLO contributions for decay modes of $\pi\pi$, $K\pi$, and $K \bar{K}$ final states.}
	\begin{threeparttable}
		\setlength{\tabcolsep}{4.0mm}{
			\begin{tabular}{c|c|cHH|c|c}
				\hline
				\hline
				&$\mathrm{LO_{NLOWC}}$		&  NLO  & NLO+soft$^*$  & NLO+soft$^a$  & NLO+soft 	& Data   \\ \hline
				Br($B^0\rightarrow \pi^+ \pi^-$) &3.90	& 4.82 &5.24 & 5.28	& $5.14\pm0.69^{+0.16+0.28}_{-0.25-0.24}$ 	&$5.12\pm0.19$   \\
				Br($B^+\rightarrow \pi^+ \pi^0$)    &3.59	& 3.24 & 5.70 & 5.70  & $5.63\pm0.53^{+0.12+0.19}_{-0.23-0.20}$  &   $5.5\pm0.4$     \\
				Br($B^0\rightarrow \pi^0 \pi^0$) &0.36	& 0.12 & 1.42 & 1.35  & $1.41\pm0.35^{+0.04+0.12}_{-0.06-0.17}$	  &$1.59\pm0.26$   \\
				Br($B^+\rightarrow K^0 \pi^+$)& 13.4	&	13.8  & 22.5 & 23.4 & $24.4\pm3.9^{+0.6+0.9}_{-0.9-0.8}$& $23.7\pm0.8$   \\
				Br($B^+\rightarrow K^+ \pi^0$)    & 9.0 &  8.4  &  12.6 & 12.5	&  $12.7\pm1.8^{+0.3+0.3}_{-0.5-0.3}$& $12.9\pm0.5$     \\
				Br($B^0\rightarrow K^+ \pi^{-}$) &13.7 & 13.2 & 23.9 &21.9 & $21.6\pm3.6^{+0.5+0.7}_{-0.7-0.6}$& $ 19.6\pm0.5$   \\
				Br($B^0\rightarrow K^0 \pi^{0}$)   & 4.9 & 5.2 &  10.5 & 10.0 &  $10.1\pm1.9^{+0.3+0.5}_{-0.4-0.3}$ & $9.9\pm0.5$      \\
				Br($B^+\rightarrow K^+ \bar{K}^0$) & 0.92	&	0.66  & 1.03 & - & $1.26\pm0.44^{+0.03+0.09}_{-0.05-0.06}$ & $1.31\pm0.17$   \\
				Br($B^0\rightarrow K^0 \bar{K}^0$)    & 0.98 &  0.68  &  1.18 & -	&   $1.34\pm0.48^{+0.04+0.11}_{-0.05-0.08}$   &$1.21\pm0.16$  \\
				Br($B^0\rightarrow K^+ K^{-}$) &0.034 & 0.034 & 0.053 &- & $0.052\pm0.013^{+0.003+0.008}_{-0.002-0.004}$& $ 0.078\pm0.015$   \\
				\hline
				\hline
				$A_{CP}$($B^0\rightarrow \pi^+ \pi^-$) &0.27 	 & 0.16 & 0.39 & 	0.40  &$0.31\pm0.03^{+0.01+0.04}_{-0.00-0.04}$ 	&$0.32\pm0.04$    \\
				$A_{CP}$($B^+\rightarrow \pi^+ \pi^0$)   & 0.00   &0.00	&0.00  &0.00   &	$0.0006\pm0.0010^{+0.0001+0.0001}_{-0.0000-0.0001}$ 	 & $0.03\pm0.04$   \\
				$A_{CP}$($B^0\rightarrow \pi^0 \pi^0$) & -0.60 	& 0.30 & 0.49	& 0.52  & $0.45\pm0.06^{+0.01+0.07}_{-0.01-0.05}$ 	 &$0.33\pm0.22$  \\
				$A_{CP}$($B^+\rightarrow K^0 \pi^+$) & -0.004  &	0.010 & 0.010&   0.012 &   $0.0106\pm0.0011^{+0.0002+0.0009}_{-0.0001-0.0010}$  &	$-0.017\pm0.016$    \\
				$A_{CP}$($B^+\rightarrow K^+ \pi^0$)   & -0.15 & -0.039 	& 0.048 & 0.078   &  $0.067\pm0.027^{+0.001+0.013}_{-0.001-0.014}$  & $0.037\pm0.021$   \\
				$A_{CP}$($B^0\rightarrow K^+ \pi^{-}$) & -0.175 & -0.107  &-0.145		&-0.082  &$-0.080\pm0.028^{+0.002+0.018}_{-0.003-0.017}$  & $-0.083\pm 0.004$  \\
				$A_{CP}$($B^0\rightarrow K^0 \pi^{0}$)   & 0.018 & -0.036	& -0.16 & -0.15  & $-0.13\pm0.04^{+0.01+0.01}_{-0.00-0.01}$   &  $0.00\pm 0.13$     \\
				$A_{CP}$($B^+\rightarrow K^+ \bar{K}^0$) & 0.07  &	0.12 & 0.02&   - & $-0.02\pm0.04^{+0.00+0.03}_{-0.01-0.02}$ & 	$0.04\pm0.014$    \\
				$A_{CP}$($B^0\rightarrow K^0 \bar{K}^0$)   & 0.00 & 0.05 	& -0.002 & -   & $-0.04\pm0.04^{+0.00+0.01}_{-0.00-0.01}$  &$-0.58^{+0.73}_{-0.66}$   \\
				$A_{CP}$($B^0\rightarrow K^+ K^{-}$) & 0.001 & 0.26  &-0.26	&-  &$-0.30\pm0.11^{+0.03+0.10}_{-0.06-0.11}$  & $-$ \\
				\hline
				\hline
		\end{tabular}  }		
	\end{threeparttable}
	
\end{table*}

	
	\begin{table*}[htb]
	\renewcommand\arraystretch{1.5}
	\caption{\label{table-BRCP-II} Branching ratio ($\times 10^{-6}$) and direct $CP$ violation with NLO contributions with decay modes involving $\eta$ and $\eta'$ mesons.}
	\begin{threeparttable}
		\setlength{\tabcolsep}{4.0mm}{
			\begin{tabular}{c|c|c|cHH|c|c}
				\hline
				\hline
				&$\mathrm{LO_{NLOWC}}$	& NLO	& NLO+gg  & NLO+soft$^*$  	& NLO+soft$^a$	 & NLO+soft &  Data   \\ \hline
				Br($B^0\rightarrow \pi^0 \eta$) &0.09	&	0.18 & 0.19 & 0.42	& 0.41 &$0.42\pm0.07^{+0.01+0.02}_{-0.01-0.02}$  & $0.41\pm0.17$   \\
				Br($B^+\rightarrow \pi^+ \eta$)   &0.97 & 1.49	& 1.52 & 5.92  &5.27	& $4.51\pm0.71^{+0.09+0.10}_{-0.11-0.09}$   & $4.02\pm0.27$     \\
				Br($B^0\rightarrow \pi^0 \eta^{'}$) &0.04 &0.14 & 0.16 & 0.23 & 0.35 & $0.85\pm0.16^{+0.01+0.03}_{-0.01-0.03}$ 	& $1.2\pm0.6$   \\
				Br($B^+\rightarrow \pi^+ \eta^{'}$)    &0.50 & 0.60 & 0.74 & 4.19 & 3.95	& $3.25\pm0.48^{+0.03+0.08}_{-0.05-0.08}$ &  $2.7\pm0.9$      \\
				Br($B^0\rightarrow K^0 \eta$) & 3.29	&	3.76 & 3.69 & 1.46 & 1.32 & $1.29\pm0.51^{+0.08+0.11}_{-0.10-0.11}$ &$1.23_{-0.24}^{+0.27}$   \\
				Br($B^+\rightarrow K^+ \eta$)    & 3.68 &  4.51	& 4.45  & 2.23 & 2.20 & $2.06\pm0.85^{+0.10+0.14}_{-0.14-0.14}$&  $2.4\pm0.4$     \\
				Br($B^0\rightarrow K^0 \eta^{'}$) & 22.4 & 30.4 & 32.6  & 64.4 & 64.1 & $ 66.6\pm21.8^{+1.2+4.3}_{-1.4-4.1}$ & $ 66\pm4$   \\
				Br($B^+\rightarrow K^+ \eta^{'}$)    & 24.8 & 33.6 & 36.0 & 71.7  & 70.9 & $ 70.0\pm23.2^{+1.2+5.0}_{-1.5-4.7}$& $70.4\pm2.5$      \\
				\hline
				\hline
				$A_{CP}$($B^0\rightarrow \pi^0 \eta$) &0.42  &	-0.06 &  -0.07 & -0.49 & -0.59 &$-0.98\pm0.03^{+0.01+0.02}_{-0.01-0.02}$ 	&-    \\
				$A_{CP}$($B^+\rightarrow \pi^+ \eta$)   & 0.40 & 0.08	& 0.06 &-0.13 &-0.14	& $-0.12\pm0.11^{+0.01+0.04}_{-0.01-0.04}$  & $-0.14\pm0.07$   \\
				$A_{CP}$($B^0\rightarrow \pi^0 \eta^{'}$) & 0.43 & 0.02	& -0.04	& 0.11 & 0.02 & $-0.42\pm0.07^{+0.01+0.06}_{-0.01-0.06}$  &-  \\
				$A_{CP}$($B^+\rightarrow \pi^+ \eta^{'}$)   & 0.51 & 0.47	& 0.36 & -0.08 & -0.09 & $0.02\pm0.11^{+0.01+0.04}_{-0.01-0.03}$ & $0.06\pm0.16$     \\
				$A_{CP}$($B^0\rightarrow K^0 \eta$) & -0.001  &	-0.05 &  -0.05  & -0.35 & -0.26 &  $-0.31\pm0.16^{+0.02+0.03}_{-0.01-0.03}$ &	-    \\
				$A_{CP}$($B^+\rightarrow K^+ \eta$)   & 0.05 & -0.05	& -0.06	& -0.34 & -0.34  & $-0.30\pm0.20^{+0.01+0.04}_{-0.01-0.04}$ & $-0.37\pm0.08$   \\
				$A_{CP}$($B^0\rightarrow K^0 \eta^{'}$) & -0.005 & 0.02	& 0.02 & 0.07 &0.06 &$0.04\pm 0.02^{+0.00+0.00}_{-0.00-0.00}$	 & $0.06\pm 0.04$  \\
				$A_{CP}$($B^+\rightarrow K^+ \eta^{'}$)   & -0.06 & -0.03	& -0.02	&-0.001 & -0.006 & $-0.003\pm 0.016^{+0.001+0.003}_{-0.001-0.003}$&  $0.004\pm 0.011$     \\
				\hline
				\hline
		\end{tabular}  }
	\end{threeparttable}	
\end{table*}
	
For the color-octet parameters and the meson pair production form factors, they cannot be calculated perturbatively in QCD because of their nonperturbative property. These parameters are treated as phenomenological parameters in this work, which can be constrained by experimental data. There are plenty of experimental data on the branching ratios and $CP$ violations for $B\to PP$ decays up to now \cite{PDG2022}, which can be used to determine these nonperturbative parameters. The color-octet parameters and the meson pair production form factors can be expressed in terms of SU(3) symmetry and symmetry-breaking parameters, which are given in Tables \ref{Y8cab} and \ref{S-Form}. We find the fitted numerical results for these parameters that can well explain the experimental data are

\begin{eqnarray}
	&&c_0^a=(0.182\pm0.015) \mathrm{Exp}[(-0.60\pm0.02)\pi i],  \nonumber \\
	&&R_1^{1a}=(0.89\pm0.03) \mathrm{Exp}[(-0.76\pm0.02)\pi i],  \nonumber \\
	&&R_2^{1a}=(0.33\pm0.03) \mathrm{Exp}[(0.52\pm0.06)\pi i], \label{input-p-a} \\
    &&R_2^{2a}=(0.45\pm0.04) \mathrm{Exp}[(0.14\pm0.08)\pi i],  \nonumber \\
    &&R_2^{4a}=(0.29\pm0.03) \mathrm{Exp}[(-0.47\pm0.08)\pi i].  \nonumber
\end{eqnarray}
\begin{eqnarray}
	&&c_0^b=(0.084\pm0.008) \mathrm{Exp}[(-0.57\pm0.02)\pi i],  \nonumber \\
	&&R_1^{1b}=(0.87\pm0.02) \mathrm{Exp}[(0.34\pm0.02)\pi i], \nonumber \\
    &&R_2^{1b}=(0.37\pm0.04) \mathrm{Exp}[(-0.57\pm0.11)\pi i], \label{input-p-b} \\
    &&R_2^{2b}=(0.26\pm0.03) \mathrm{Exp}[(-0.36\pm0.08)\pi i],  \nonumber \\
    &&R_2^{4b}=(0.37\pm0.03) \mathrm{Exp}[(0.28\pm0.14)\pi i].  \nonumber
\end{eqnarray}
\begin{eqnarray}	
	&&c_0^c=(0.56\pm0.04) \mathrm{Exp}[(-0.68\pm0.02)\pi i],  \nonumber \\
	&&R_1^{1c}=(0.32\pm0.02) \mathrm{Exp}[(0.72\pm0.07)\pi i],  \nonumber \\
	&&R_1^{2c}=(0.57\pm0.07) \mathrm{Exp}[(0.92\pm0.08)\pi i],  \label{input-p-c}\\
	&&R_2^{1c}=(0.44\pm0.03) \mathrm{Exp}[(-0.43\pm0.10)\pi i].  \nonumber
\end{eqnarray}

The comparison of the theoretical results about the branching ratios and $CP$ violations with experimental data is presented in Tables \ref{table-BRCP-I} and \ref{table-BRCP-II}, where the column ``$\mathrm{LO_{NLOWC}}$" shows the leading-order contributions in QCD but with NLO Wilson coefficients being used, the column ``NLO" shows the main NLO contribution in QCD with the NLO Wilson coefficient used, ``NLO+$gg$" shows the NLO contribution in QCD plus the contribution of $g^*g^*\eta^{(\prime)}$ effective coupling, and "NLO+soft" shows both the contributions of NLO in QCD, the soft form factors and the color-octet contributions included, where the first errors are caused by the uncertainties of soft form factors and color-octet parameters, the second and third errors are caused by the uncertainties of the parameters in the wave functions of $B$ and light mesons, respectively. The difference between NLO and $\mathrm{LO_{NLOWC}}$ shows the NLO corrections. Tables \ref{table-BRCP-I} and \ref{table-BRCP-II} show that the NLO corrections to branching ratios are at most up to 10\% to 20\% for tree-level non-color-suppressed decays. For most decay modes, the NLO corrections are only at the order of a few percent. Only for the few decay modes where the tree-level contributions are suppressed, are the NLO contributions relatively large. The contribution of the $g^*g^*\eta^{(\prime)}$ effective coupling is generally small (see Table \ref{table-BRCP-II}). Only after including the contributions of the soft form factors and color-octet contributions, can the theoretical results be consistent with experimental data. Tables \ref{table-BRCP-I} and \ref{table-BRCP-II} show that our results are all in good agreement with the data for both branching ratios and $CP$ violations. Therefore, the $\pi\pi$ and $K\pi$ puzzles are solved in a systematic way.

It is pointed out in Ref. \cite{Li-Mishima2006} that the experimental data of $B\to \rho\rho$ decays have seriously constrained the possibility of resolving the $B\to \pi\pi$ puzzle in the theoretical approaches, such as the PQCD and QCDF approaches, which are based on the factorization theorem in QCD. The predictions of NLO PQCD for the branching ratios of $B^0\to \rho^\mp\rho^\pm$ and $B^\pm\to \rho^\pm\rho^0$ are consistent with experimental data, and the branching ratio of $B^0\to \rho^0\rho^0$ has been close to the experimental upper limit, while the prediction for the branching ratio of $B^0\to \pi^0\pi^0$ is still much smaller than experimental data. The QCDF with the inclusion of the NLO jet function from the soft-collinear effective theory, however, can enhance the branching of $B^0\to \pi^0\pi^0$ sufficiently. it exceeds the upper limit of the branching ratio for $B^0\to \rho^0\rho^0$ decay mode \cite{Li-Mishima2006}. Then a question is whether the present approach in this work can predict the branching ratio of $B^0\to \rho^0\rho^0$ in accord with the experimental upper limit while resolving the puzzles of $B\to \pi\pi$ and $K\pi$ decays simultaneously. In the approaches based on the factorization theorem, the meson wave functions are universal, and the short-distance contributions, such as the vertex corrections, the quark loop, and the magnetic penguin are similar for different final states in the decay modes. Therefore, the constraint from the data of the branching ratio of $B^0\to \rho^0\rho^0$ decay is serious. In the present approach, the introduction of the soft cutoff scale and the inclusion of the contributions of the soft form factors, and especially the color-octet contributions, changed the contribution structure of PQCD in the earlier stage. The color-octet contribution can be final-state dependent, because it is essentially long-distance contribution. In the present work for $B\to PP$ decays, we find a set of universal parameters for the color-octet contributions for the final mesons within one SU(3) nonet by considering the SU(3) flavor symmetry and its symmetry breaking. For $B\to PV$ and $VV$ decays, where $V$ stands for vector meson, the parameters for the color-octet contributions may be slightly different from that for $PP$ final states. It may depend on the SU(3) flavor nonet of vector mesons. The serious constraint from the experimental upper limit for the branching ratio of the $B^0\to \rho^0\rho^0$ decay can be evaded by different long-distance interactions. It is indeed interesting to see if our present approach can predict if the branching ratios and $CP$ violations are consistent with experimental data with the nonperturbative inputs in the reasonable parameter space. As a preliminary investigation, we tried some values for soft parameters for $B\to\rho\rho$ decays to check what the output for the branching ratios and $CP$ violations are for these decays. Table \ref{table-BRCP-III} is for the results of the branching ratios and $CP$ violations for $B\to\rho\rho$ decays with the color-octet parameters $Y^8_F=0.196{\rm Exp}(-0.453\pi i)$ and $Y^8_M=0.201{\rm Exp}(-0.367\pi i)$, and the production form factor $F_+^{\rho\rho}=0.290{\rm Exp}(-0.826\pi i)$. It shows that both the branching ratios and $CP$ violations for $B\to\rho\rho$ decays are consistent with experimental data. The soft parameters used here can be compared with that used for $B\to PP$ decays. Table \ref{Y8cab2} is for the soft parameters for each decay mode of $B\to PP$ decays, which are obtained by using Eqs. (\ref{input-p-a})$\--$(\ref{input-p-c}) and Tables \ref{Y8cab} and \ref{S-Form}. From Table \ref{Y8cab2}, we can see that the soft parameters $Y^8_{F,M}$ and $F_+^{\rho\rho}$ used for $B\to\rho\rho$ decays are within the range of the relevant parameters for $B\to PP$ decays. Therefore, it is convincing that the present approach can explain the experimental data of $B\to \rho\rho$ simultaneously. The detailed study for these decays will be given elsewhere in the near future. 

\begin{table}[htb]
	\renewcommand\arraystretch{1.5}
	\caption{\label{table-BRCP-III} Branching ratio ($\times 10^{-6}$) and direct $CP$ violation of $B\to\rho\rho$ decays with soft parameters $F_+^{\rho\rho}=0.290{\rm Exp}(-0.826\pi i)$, $Y^8_F=0.196{\rm Exp}(-0.453\pi i)$, and $Y^8_M=0.201{\rm Exp}(-0.367\pi i)$.}
	\begin{threeparttable}
			\begin{tabular}{c|c|c|c|c}
				\hline
				\hline
									&$\mathrm{LO_{NLOWC}}$	& NLO		& NLO+soft	&  Data   \\
				\hline
				Br($B^+\to\rho^+\rho^0$) 		& $6.9 $	& $6.2 $	& $22.2$	& $24.0\pm 1.9$ 	\\
				Br($B^0\to\rho^+\rho^-$) 		& $9.8 $	& $11.1$	& $27.7$	& $27.7\pm 1.9$ 	\\
				Br($B^0\to\rho^0\rho^0$) 		& $0.31$	& $0.07$	& $1.08$	& $0.96\pm 0.15$	\\
				\hline
				\hline
				$A_{CP}(B^+\to\rho^+\rho^0)$	&$ 0.00$	&$ 0.00$	&$ 0.00$	&$-0.05\pm 0.05$	\\
				$A_{CP}(B^0\to\rho^+\rho^-)$	&$-0.03$	&$-0.08$	&$-0.09$	&$0.00\pm 0.09$   	\\
				$A_{CP}(B^0\to\rho^0\rho^0)$	&$ 0.21$	&$ 0.79$	&$ 0.60$	&$-0.2\pm 0.9 $ 	\\
				\hline
				\hline
		\end{tabular}
	\end{threeparttable}	
\end{table}

\begin{table*}[htb]
\renewcommand\arraystretch{1.5}
\caption{\label{Y8cab2} The soft parameters of different decay modes of $B\to PP$ decays.}
	\begin{threeparttable}
		\setlength{\tabcolsep}{7mm}{
			\begin{tabular}{c|c|c|c}
				\hline
				\hline
				&$Y^8_F$ &  $Y^8_M$   &  $ F_+^{M_1 M_2}$  \\ \hline
				$\pi \pi(\eta_q)$    & 0.182 $e^{-0.60\pi i}$ & 0.085 $e^{-0.57pi i}$ &  0.319$e^{-0.68\pi i}$ \\
				$K \pi(\eta_q)$   &0.135$e^{-0.74\pi i}$   &  0.108$e^{-0.48\pi i}$ &  0.265$e^{-0.58\pi i}$	   \\
				$\pi \eta_s$ 	  & 0.154$e^{-0.96\pi i}$  &  0.151$e^{-0.40\pi i}$ & 0 \\
				$K \eta_s$      & 0.124$e^{0.99\pi i}$  &  0.162$e^{-0.43\pi i}$  &  0.107$e^{-0.71\pi i}$   \\
				$K \bar{K}$       &  0.139$e^{-0.84\pi i}$  & 0.137$e^{-0.44\pi i}$   &  0.149$e^{-0.58\pi i}$  \\
				\hline
				\hline
		\end{tabular}  }
	\end{threeparttable}
\end{table*}

\section{Summary}

We study $B\to PP$ decays in the modified PQCD approach, where the wave function of $B$ meson obtained by solving the wave equation in the QCD inspired relativistic potential model is used. A critical soft momentum cutoff scale $\mu_c$ is introduced. For the contributions with the scale $\mu >\mu_c$, the decay amplitudes are calculated with the PQCD approach. For the contributions in the region of lower scale $\mu <\mu_c$, soft form factors are introduced. The soft contributions are absorbed into these soft form factors. In addition, the color-octet states for the final mesons are considered. The color-octet contributions are included, which are essentially of long-distance property. With these soft contributions included, the branching ratios and $CP$ violations are calculated. By selecting reasonable values for the input parameters, the results of our theoretical calculation for all the $B\to PP$ decay modes are consistent with experimental data.

\vspace{0.5cm}
\acknowledgments
This work is supported in part by the National Natural Science Foundation of China under
Contracts No. 12275139 and No. 11875168.

\appendix{}
\section{\label{a}SUDAKOV FACTOR AND ULTRAVIOLET LOGARITHMS IN QCD}

The threshold factor $S_t(x)$ can be parametrized as \cite{lihn2002}
\begin{equation}
  S_t(x)=\frac{2^{1+2c}\Gamma(3/2+c)}{\sqrt{\pi}\Gamma(1+c)}[x(1-x)]^c,
\end{equation}
with $c=0.3$.

The exponentials $\exp[-S_{B}(\mu)]$,  $\exp[-S_{M_1}(\mu)]$, and $\exp[-S_{M_2}(\mu)]$ are the Sudakov factor and the relevant single ultraviolet logarithms associated with the heavy and light mesons. The exponents are \begin{equation}
S_B(\mu) = s(x,b,m_B)-\frac{1}{\beta_1}\ln \frac{\ln (\mu/\Lambda_{\mbox{QCD}})}
         {\ln (1/(b\Lambda_{\mbox{QCD}}))}
\end{equation}
\begin{eqnarray}
S_{M_1}(\mu) &=& s(x_1,b_1,m_B)+s(1-x_1,b_1,m_B)\nonumber\\
&&\;\;-\frac{1}{\beta_1}\ln \frac{\ln (\mu/\Lambda_{\mbox{QCD}})}
         {\ln (1/(b_1\Lambda_{\mbox{QCD}}))}
\end{eqnarray}
\begin{eqnarray}
S_{M_2}(\mu) &=& s(x_2,b_2,m_B)+s(1-x_2,b_2,m_B)\nonumber\\
&&\;\;-\frac{1}{\beta_1}\ln \frac{\ln (\mu/\Lambda_{\mbox{QCD}})}
         {\ln (1/(b_2\Lambda_{\mbox{QCD}}))}
\end{eqnarray}
\vspace{0.6cm}

The exponent $S(x,b,Q)$ up to next-to-leading order in QCD is \cite{Li1995}
\begin{widetext}
\begin{eqnarray}
&& s(x,b,Q)=\frac{A^{(1)}}{2\beta_{1}}\hat{q}\ln\left(\frac{\hat{q}}
{\hat{b}}\right)-
\frac{A^{(1)}}{2\beta_{1}}\left(\hat{q}-\hat{b}\right)+
\frac{A^{(2)}}{4\beta_{1}^{2}}\left(\frac{\hat{q}}{\hat{b}}-1\right)
-\left[\frac{A^{(2)}}{4\beta_{1}^{2}}-\frac{A^{(1)}}{4\beta_{1}}
\ln\left(\frac{e^{2\gamma_E-1}}{2}\right)\right]
\ln\left(\frac{\hat{q}}{\hat{b}}\right)
\nonumber \\
&&+\frac{A^{(1)}\beta_{2}}{4\beta_{1}^{3}}\hat{q}\left[
\frac{\ln(2\hat{q})+1}{\hat{q}}-\frac{\ln(2\hat{b})+1}{\hat{b}}\right]
+\frac{A^{(1)}\beta_{2}}{8\beta_{1}^{3}}\left[
\ln^{2}(2\hat{q})-\ln^{2}(2\hat{b})\right]
\nonumber \\
&&+\frac{A^{(1)}\beta_{2}}{8\beta_{1}^{3}}
\ln\left(\frac{e^{2\gamma_E-1}}{2}\right)\left[
\frac{\ln(2\hat{q})+1}{\hat{q}}-\frac{\ln(2\hat{b})+1}{\hat{b}}\right]
-\frac{A^{(1)}\beta_{2}}{16\beta_{1}^{4}}\left[
\frac{2\ln(2\hat{q})+3}{\hat{q}}-\frac{2\ln(2\hat{b})+3}{\hat{b}}\right]
\nonumber \\
& &-\frac{A^{(1)}\beta_{2}}{16\beta_{1}^{4}}
\frac{\hat{q}-\hat{b}}{\hat{b}^2}\left[2\ln(2\hat{b})+1\right]
+\frac{A^{(2)}\beta_{2}^2}{432\beta_{1}^{6}}
\frac{\hat{q}-\hat{b}}{\hat{b}^3}
\left[9\ln^2(2\hat{b})+6\ln(2\hat{b})+2\right]
\nonumber \\
&& +\frac{A^{(2)}\beta_{2}^2}{1728\beta_{1}^{6}}\left[
\frac{18\ln^2(2\hat{q})+30\ln(2\hat{q})+19}{\hat{q}^2}
-\frac{18\ln^2(2\hat{b})+30\ln(2\hat{b})+19}{\hat{b}^2}\right]
\label{sss}
\end{eqnarray}
\end{widetext}
where $\hat q$ and $\hat b$ are defined by
\begin{equation}
{\hat q} \equiv  {\rm ln}\left(xQ/(\sqrt 2\Lambda_{QCD})\right),~
{\hat b} \equiv  {\rm ln}(1/b\Lambda_{QCD})
\end{equation}
The coefficients $\beta_{i}$ and $A^{(i)}$ are
\begin{eqnarray}
& &\beta_{1}=\frac{33-2n_{f}}{12}\;,\;\;\;\beta_{2}=\frac{153-19n_{f}}{24}\; ,
A^{(1)}=\frac{4}{3}\;,
\nonumber \\
& & A^{(2)}=\frac{67}{9}-\frac{\pi^{2}}{3}-\frac{10}{27}n_
{f}+\frac{8}{3}\beta_{1}\ln\left(\frac{e^{\gamma_E}}{2}\right)\;
\end{eqnarray}
and $\gamma_E$ is the Euler constant.

\section{\label{b}LIGHT MESON DISTRIBUTION AMPLITUDES}
The transverse-momentum-dependent light meson distribution amplitudes are
$\phi_M(x,k_{q\perp})$, $\phi_P^M(x,k_{q\perp})$, and $\phi_\sigma^M(x,k_{q\perp})$,
where $M$ represents pion, kaon, or $\eta_{q,s}$ mesons. The transverse-momentum dependence
is assumed to be a Gaussian form and appears as a factorized part from the longitudinal wave functions.
Transformed into $b$ space, the distribution amplitudes can be written as \cite{wy2002}
\begin{equation}
    \phi(x,b)=\phi(x)\exp\left(-\frac{b^2}{4\beta^2}\right).
\end{equation}
Here, we denote the $b$-space distribution amplitudes as $\phi_M(x,b)$, $\phi_P^M(x,b)$, and $\phi_\sigma^M(x,b)$.
As discussed previously in Ref. \cite{wang-yang2023} (see also Refs. \cite{wy2002} and \cite{JK93}),
we adopt $\beta=4.0 \;\mathrm{GeV}^{-1} $ for the wave functions of pion, kaon, and $\eta_{q,s}$ mesons.
The expressions for the twist-2 and twist-3 distribution amplitudes are given by \cite{Ball-Braun2006}
\begin{equation}
  \label{eq:phi}
  \phi_M(x)=6x(1-x)\biggl[1+a_1^M C_1^{3/2}(t)+a_2^M C_2^{3/2}(t)\biggr],
\end{equation}
\begin{equation}
  \label{eq:phip}
  \begin{split}
    \phi_P^M(x)&=1+a_{0P}^M+a_{1P}^MC_1^{1/2}(t)+a_{2P}^MC_2^{1/2}(t) \\
      &\quad+a_{3P}^MC_3^{1/2}(t) +a_{4P}^MC_4^{1/2}(t) \\
      &\quad+b_{1P}^M\ln(x)+b_{2P}^M\ln(1-x),\\
  \end{split}
\end{equation}
\begin{equation}
  \label{eq:phis}
  \begin{split}
    \phi_\sigma^M(x)&=6x(1-x)\biggl[1+a_{0\sigma}^M+a_{1\sigma}^MC_1^{3/2}(t) \\
      &\quad+a_{2\sigma}^MC_2^{3/2}(t)+a_{3\sigma}^MC_3^{3/2}(t)\biggr] \\
      &\quad+9x(1-x)\biggl[b_{1\sigma}^M\ln(x)+b_{2\sigma}^M\ln(1-x)\biggr],\\
  \end{split}
\end{equation}
where $t$ is defined as $t=2x-1$.
These $C$ functions are Gegenbauer polynomials.
The coefficients appearing in Eqs.~\eqref{eq:phi}--\eqref{eq:phis},
with $a_{i(P,\sigma)}^M$ for $i = 1,2,3,4$ and $b_{j(P,\sigma)}^M$ for $j =1,2$,
have the following values:
\begin{equation}
  \begin{split}
    &a_1^\pi=0, \quad a_2^\pi=0.25\pm 0.15, \\
    &a_{0P}^\pi=0.048\pm 0.017, \quad a_{2P}^\pi=0.62\pm 0.21,\\
    &a_{4P}^\pi=0.089\pm 0.071, \quad a_{1P}^\pi=a_{3P}^\pi=0, \\
    &b_{1P}^\pi=b_{2P}^\pi=0.024\pm 0.009, \\
    &a_{0\sigma}^\pi=0.034\pm 0.014,\quad a_{2\sigma}^\pi=0.12\pm 0.05, \\
    &a_{1\sigma}^\pi=a_{3\sigma}^\pi=0, \quad b_{1\sigma}^\pi=b_{2\sigma}^\pi=0.016\pm 0.006, \\
\end{split}
\end{equation}
for the pion,
\begin{eqnarray}
    &&a_1^K=0.06\pm 0.03, \quad a_2^K=0.25\pm 0.15, \nonumber \\
    &&a_{0P}^K=0.59\pm 0.24, \quad a_{1P}^K=-0.52\pm 0.32,\nonumber \\
	&&a_{2P}^K=0.79\pm 0.36, \quad a_{3P}^K=0.18\pm 0.20,\nonumber \\
    &&a_{4P}^K=0.06\pm 0.05, \nonumber\\
    &&b_{1P}^K=0.54\pm 0.23, \quad b_{2P}^K=0.05\pm 0.02,\nonumber \\
    &&a_{0\sigma}^K=0.41\pm 0.20, \quad a_{1\sigma}^K=-0.12\pm 0.09, \nonumber\\
    &&a_{2\sigma}^K=0.12\pm 0.06, \quad a_{3\sigma}^K=0.03\pm 0.02, \nonumber\\
    &&b_{1\sigma}^K=0.36\pm 0.15, \quad b_{2\sigma}^K=0.03\pm 0.01, 
\end{eqnarray}
for the kaon,
\begin{eqnarray}
    &&a_1^{\eta_q}=0, \quad a_2^{\eta_q}=0.25\pm 0.15,\nonumber \\
    &&a_{0P}^{\eta_q}=0.079\pm 0.028, \quad a_{2P}^{\eta_q}=0.95\pm 0.33,\nonumber\\
    &&a_{4P}^{\eta_q}=0.14\pm 0.11, \quad a_{1P}^{\eta_q}=a_{3P}^{\eta_q}=0, \nonumber\\
    &&b_{1P}^{\eta_q}=b_{2P}^{\eta_q}=0.039\pm 0.014, \nonumber\\
    &&a_{0\sigma}^{\eta_q}=0.055\pm 0.024,\quad a_{2\sigma}^{\eta_q}=0.18\pm 0.07, \nonumber\\
    &&a_{1\sigma}^{\eta_q}=a_{3\sigma}^{\eta_q}=0, \quad b_{1\sigma}^{\eta_q}=b_{2\sigma}^{\eta_q}=0.026\pm 0.009, \nonumber\\
\end{eqnarray}
for the $\eta_q$ meson, and
\begin{equation}
  \begin{split}
    &a_1^{\eta_s}=0, \quad a_2^{\eta_s}=0.25\pm 0.15, \\
    &a_{0P}^{\eta_s}=1.13\pm 0.41, \quad a_{2P}^{\eta_s}=0.99\pm 0.48,\\
    &a_{4P}^{\eta_s}=0.06\pm 0.05, \quad a_{1P}^{\eta_s}=a_{3P}^{\eta_s}=0, \\
    &b_{1P}^{\eta_s}=b_{2P}^{\eta_s}=0.56\pm 0.20, \\
	\end{split}
\end{equation}
\vspace{1em}
\begin{equation}
	\begin{split}
    &a_{0\sigma}^{\eta_s}=0.79\pm 0.34,\quad a_{2\sigma}^{\eta_s}=0.14\pm 0.07, \\
    &a_{1\sigma}^{\eta_s}=a_{3\sigma}^{\eta_s}=0, \quad b_{1\sigma}^{\eta_s}=b_{2\sigma}^{\eta_s}=0.38\pm 0.14, \\
\end{split}
\end{equation}
for the $\eta_s$ meson.
The parameters listed above are all determined at the renormalization scale of $\mu=1.0~\mathrm{GeV}$.
It is worth noting that, considering the similarity in quark composition between $\eta_{q,s}$ meson and pion,
we employ the same expressions for $\eta_{q,s}$ meson parameters as for the pion,
with appropriate substitutions made only for parts involving meson masses, quark masses, and decay constants.
The Gegenbauer polynomials are given by
\begin{equation}
  \begin{split}
    &C_1^{1/2}(t)=t,\\
    &C_2^{1/2}(t)=\frac{1}{2}\left(3t^2-1\right), \\
    &C_3^{1/2}(t)=\frac{t}{2}\left(5t^2-3\right), \\
    &C_4^{1/2}(t)=\frac{1}{8}\left(35t^4-30t^2+3\right), \\
  \end{split}
\end{equation}
and
\begin{equation}
  \begin{split}
    &C_1^{3/2}(t)=3t, \\
    &C_2^{3/2}(t)=\frac{3}{2}\left(5t^2-1\right), \\
    &C_3^{3/2}(t)=\frac{5}{2}t\left(7t^2-3\right), \\
    &C_4^{3/2}(t)=\frac{15}{8}\left(21t^4-14t^2+1\right). \\
  \end{split}
\end{equation}

\end{document}